\documentclass[11pt]{article}

 \usepackage{graphicx}  
 \usepackage{amsmath}
 \usepackage{amssymb}
 \usepackage[mathscr]{eucal}
 \usepackage{fancyhdr}
\usepackage{mathpazo}

\def\bfa{{\bf a}}
\def\bfA{{\bf A}}

\def\bfb{{\bf b}}
\def\bfC{{\bf C}}

\def\bff{{\bf f}}
\def\bfg{{\bf g}}
\def\bfgamma{{\boldsymbol{\gamma}}}

\def\bfH{{\bf H}}
\def\bfI{{\bf I}}
\def\bfL{{\bf L}}
\def\bflambda{{\boldsymbol{\lambda}}}

\def\bfomega{{\boldsymbol{\omega}}}
\def\bfOmega{{\boldsymbol{\Omega}}}
\def\bfone{{\boldsymbol{1}}}
\def\bfP{{\bf P}}

\def\bfpi{{\boldsymbol{\pi}}}
\def\bfPi{{\boldsymbol{\Pi}}}
\def\PointP{{\mathscr{P}}}

\def\bfsigma{{\boldsymbol{\sigma}}}
\def\bfSigma{{\boldsymbol{\Sigma}}}
\def\bft{{\bf t}}
\def\bfT{{\bf T}}
\def\bftau{{\boldsymbol{\tau}}}
\def\bfW{{\bf W}}
\def\bfz{{\bf z}}
\def\bfZ{{\bf Z}}

\def\calG{{\cal G}}
\def\computed{{\rm computed}}

\def\obs{{\rm obs}}
\def\prior{{\rm prior}}
\def\SpaceA{{\mathbb{A}}}
\def\SpaceB{{\mathbb{B}}}

\title{Gravimetry, Relativity,\\
   and the Global Navigation Satellite Systems%
   \footnote{%
   Lesson delivered at the School
   {\sl Relativistic Coordinates, Reference and Positioning Systems}
   (Salamanca, January 2005).}}
   
\author{Albert Tarantola\thanks{Institut de Physique du Globe de Paris, 
albert.tarantola@ipgp.org, {\tt http://www.ipgp.jussieu.fr/$\sim$tarantola/}.}\\[3 pt]
        Ludek Klime\v{s}\thanks{Charles University, Prague, {\tt http://sw3d.cz/staff/klimes.htm}.}\\[3 pt]
Jos\'e Maria Pozo \& Bartolom\'e Coll\thanks{Observatoire de Paris, {\tt http://syrte.obspm.fr/$\sim$coll/}.}}

\begin{document}

\setcounter{page}{5}

\maketitle

\begin{abstract}
Relativity is an integral part of positioning systems, and this is taken 
into account in today's practice by applying many `relativistic corrections' 
to computations performed using concepts borrowed from Galilean physics. 
A different, fully relativistic paradigm can be developed for operating a 
positioning system. This implies some fundamental changes. 
For instance, the basic coordinates are four times (with a symmetric meaning, 
not three space coordinate and one time coordinate) and the satellites \emph{must}
have cross-link capabilities. Gravitation must, of course,
be taken into account, but not using the Newtonian theory:
the gravitation field is, and only is, the space-time metric. This implies that
the positioning problem and the gravimetry problem can not be separated. 
An optimization theory can be developed that, because it is fully relativistic,
does not contain any `relativistic correction'. We suggest that all positioning 
satellite systems should be operated in this way. The first benefit of doing so
would be a clarification and a simplification of the theory. We also expect,
at the end, that positioning systems will provide increased positioning accuracy.
\end{abstract}

\setcounter{tocdepth}{2} \tableofcontents


\section{Introduction}

Many relativistic corrections are applied to the Global Navigation
Satellite Systems (GNSS). 
Neil Ashby presents in Physics Today (May 2002) a good account 
of how these relativistic corrections are applied, why, and which 
are their orders of magnitude. Unfortunately, it is generally 
proposed that relativity is only a correction to be applied to 
Newtonian physics. We rather believe that there is a fully relativistic 
way to understand a GNSS system, this leading to a new way of operating it.

As gravitation has to be taken into account, it is inside the framework
of general relativity that the theory must be developed.
The shift from a Newtonian viewpoint (relativistic corrections included or not)
into a relativistic framework requires some fundamental conceptual changes.
Perhaps the most important concerns the operational definition of a system
of four space-time coordinates. We reach the conclusion that there is an
(essentially unique) coordinate system that, while being consistent with 
a relativistic formulation, allows an \emph{immediate} positioning of observers 
(the traditional Minkowski coordinates \,$ \{ t,x,y,z \} $\, of flat space-time
do not allow such an immediate positioning).

These coordinates are defined as follows%
\footnote{%
Coll (1985, 2000, 2001a, 2001b, 2002), Rovelli (2001, 2002), 
Coll and Morales (1992), Blagojevi\'c et al.\ (2001), Coll and Tarantola (2003).}.
%
%
If four clocks ---having an arbitrary space-time tra\-jec\-to\-ry--- 
broadcast their proper time ---using electromagnetic signals,---
then, any observer receives, at any point along his personal 
space-time trajectory, four times, corresponding to the four
signals arriving at that space-time point. These four times,
say \,$ \{ \tau^1,\tau^2,\tau^3,\tau^4\} $\,, \emph{are},
by definition, the coordinates of the space-time point. 
One doesn't has one time coordinate and three space coordinates,
as usual, but a `symmetric' coordinate system with four time coordinates.

The space-time having been endowed with those coordinates, any observer with
a receiver may obtain (in real time) his personal trajectory.
This is true, in particular, for the four clocks themselves: 
each clock constantly receives three of the coordinates and it
defines the fourth. Therefore, \emph{each clock knows its own trajectory}
in this self-consistent coordinate system. Note that even if the clocks
are satellites around the Earth, the coordinates and the orbits are
defined without any reference to an Earth based coordinate system:
this allows to achieve maximum precision for this \emph{primary} 
reference system. Of course, for applications on the Earth's surface,
the primary coordinates must be attached to some terrestrial
coordinate system, but this is just an attachment problem that should
not interfere with the problem of defining the primary system itself.

In general relativity, the gravity field is the space-time metric. 
Should this metric be exactly known (in any coordinate system), 
the system just described would constitute an ideal positioning system
(and the components of the metric could be expressed in these coordinates). 
In practice the space-time metric (i.e., the gravity field) is not exactly 
known, and the satellite system itself has to be used to infer it. 
This article is about the problem of using a satellite system for both, 
positioning, and measuring the space-time metric.

Information on the space-time metric may come from different sources.
First, any satellite system has more than four clocks. While four of
the clocks define the coordinates, the redundant clocks can be used
to monitor the space-time metric. The considered satellites may have
more that a clock: they may have an accelerometer (this providing 
information on the space-time connection), a gradiometer (this providing 
information on the Riemann), etc. Our theory will provide seamless 
integration of positioning systems with systems designed for gravimetry.

In the ``post Newtonian'' paradigm used today for operating
positioning and gravimetry systems, the ever increasing accuracy of 
clocks makes that more and more ``relativistic effects'' have to be 
taken into account. On the contrary, the fully relativistic theory here 
developed will remain valid as long as relativity itself remains valid.

It is our feeling that when GNSS and gravimetry systems will be operated
using the principles here exposed, new experimental possibilities will
appear. One must realize that with the optical clocks being developed 
may one day have a relative accuracy of \,$ 10^{18} $\,. The possibility
that some day we may approach this accuracy for positioning immediately
suggests extraordinarily interesting applications.

These applications would simply be impossible if sticking to the present-day
paradigm. To realize how deeply nonrelativistic this paradigm is, consider
that GPS clocks are kept synchronized. In this year 2005, when we celebrate
the centenary of relativity, this sounds strange: is there anything less
relativistic than the obstination to keep synchronized a system of clocks
in relative movement?

There is one implication of the theory here developed for the Galileo
positioning system now being developed by the European Union. Our theory
requires, as a fundamental fact, that the GNSS satellites exchange signals.
The most recent GPS satellites (from the USA) do have this ``cross-link''
capability. One could, in principle, use the cross-link data
(or an ameliorated version of it) to operate the system in the way
here proposed. Unfortunately, the Galileo satellites will not have, 
to our knowledge, this cross-link capability. This is a serious limitation
that will complicate the evolution of the system towards a more precise one.

Finally, we need to write a disclaimer here. None of the algorithms
proposed below are intended to be practical. They are the simplest
algorithms that would be fully consistent with relativity theory. 
Passing from these to actually implementable algorithms will require
some developments in numerical analysis. Finally, some of the simplifying 
hypotheses made below are not necessary and are only intended to start 
with a theory that is as simple as possible.


\section{Setting of the Problem}


\subsection{Model Parameters and Observable Parameters}

Imagine that four clocks (called below the \emph{basic clocks}) 
broadcast their proper time using light signals. Any observer in space-time 
may receive, at any point along  its space-time trajectory, four times 
\,$ \{ \tau^\alpha \} = \{ \tau^1,\tau^2,\tau^3,\tau^4 \} $\,: 
these are, by definition, the (space-time) coordinates of the
space-time point where the observer is.
If the observer had his own clock, with proper time denoted 
\,$ \tau $\,, then he would know his trajectory 
\begin{equation}
\tau^\alpha \ = \ \tau^\alpha(\tau) \quad ,
\label{eq: art-001}
\end{equation}
i.e., he would know four functions (of his proper time \,$ \tau $\,).
He could, in addition, evaluate his four-velocity, and his four-acceleration
(see below).

In an actual experiment, the clocks are never ideal, and the reception of times
implies a measurement that has always attached uncertaities. We must, therefore, 
carefully distinguish between model parameters and observable parameters. 
This places the present discussion into the usual conceptual frame of inverse 
problem theory%
\footnote{%
For an introductory text, see Tarantola (2005).}.
%
%


\subsubsection{Complete Model}

For us, a \emph{complete model} consists in:
\begin{itemize}

\item
A {\bf space-time metric field}, denoted
\,$ \{ \, g_{\alpha\beta}(\tau^1,\tau^2,\tau^3,\tau^4) \, \} $\,,
and some (at least four) {\bf trajectories}, denoted
\,$ \{ \, \tau^\alpha_1(\lambda_1) \, , \, 
          \tau^\alpha_2(\lambda_2) \, , \, \dots \, \} $\,,
parameterized by some parameters 
\,$ \lambda_1 \, , \, \lambda_2 \, , \, \dots $\ \ 
Because the metric is given, 
these parameters can be converted into proper time (by integration of
the element \,$ \sqrt{g_{\alpha\beta} \, d\tau^\alpha \, d\tau^\beta} \, $\,),
so the four trajectories can always be considered to be given as a 
function of proper time, 
\,$ \{ \, \tau^\alpha_1(\tau_1) \, , \, 
          \tau^\alpha_2(\tau_2) \, , \, \dots \, \} $\,. 
These trajectories shall be the model of the trajectories of ``satellites'' 
consisting on physical clocks,
and, perhaps, accelerometers, gradiometers, and other measuring instruments.
Four of the trajectories are arbitrarily selected as `basic trajectories,
and the {\bf working coordinates} \,$ \{\tau^1,\tau^2,\tau^3,\tau^4 \} $\, 
are assumed to be linked to the metric and the four basic trajectories as follows: 
the four coordinates \,$ \{\tau^1,\tau^2,\tau^3,\tau^4 \} $\, of a space-time
point \,$ \PointP $\, are, by definition, the four emission times 
(one in each of the four trajectories) of the four light ``cones'' 
passing by the point \,$ \PointP $\,.
This is an idealization of the heuristic protocol suggested in the
introduction.

\item
Associated to each trajectory, a (typically smooth) {\bf clock drift function} 
\,$ \{ f_1(\tau_1) \, , \, f_2(\tau_2) \, , \, \dots \} $\,, describing the 
drift of the physical clock with respect to proper time 
(if \,$ z $\, is clock time, and \,$ \tau $\, is proper time, 
then \,$ z(\tau) = \tau + f(\tau) $\,).
\end{itemize}
To these fundamental parameters, we need to add another set, that are also
necessary for the prediction of observations:
\begin{itemize}

\item
A {\bf set of instants} \,$ \{ \tau_a \, , \, \tau_b \, , \, \dots \} $\, 
along each trajectory, that represent the nominal instants
when a clock is observed, a light signal is emitted (that will be received 
by some other satellite), or the instant when a measurement (acceleration, 
gradiometry, etc.) is made.

\end{itemize}

To simplify the theory, we shall assume that the space-time trajectory of
all the satellites has crossed at a given space-time point, and that all the clocks
have been synchronized (to zero time) at this point. From then on, all clocks
will follow their proper time, without any further synchronization (the drift
function just mentioned takes into account that physical clocks never exactly 
follow proper time).

We choose in this first version of the theory, not to introduce the fact that
light does not propagate in absolute vacuum. One very simple model for the upper
layers of the atmosphere would be as follows. One could assume that light
propagates in a gas that, at rest, is homogeneous and isotropic, with an
index of refraction \,$ n $\, that, in general, depends on the carrier frequency
of the signal. If the 
four-velocity of the gas is \,$ U^\alpha $\,, then Maxwell equations in the gas 
take the same form as in vacuum, excepted that one must replace the actual
(Riemannian) metric by the (Finslerian) metric (see Gordon (1923) or 
Pham Mau Quan (1957) for details)
\begin{equation}
g_{\alpha\beta} + \Big( 1-\frac{1}{n^2} \Big) \, U_\alpha \, U_\beta \quad .
\end{equation} 
When we will introduce this aspect in the theory, a complete model will also
include the field \,$ n $\, (index of refraction at different frequencies), 
and the field \,$ U^\alpha $\, (four-velocity of the gas). The a priori
information that one has on these quantities is precise enough as for hoping 
that satellite data will easily be able to refine the model (tomography of 
the ionosphere using GPS data is already a well developed topic of research).

In a physical implementation of the proposed system, the space-time trajectory 
of any satellite can 
be approximately known by just recording the time signals received from the 
four basic satelites.  But the arrival time of the signals must be \emph{measured}
and any measurement is subject to experimental uncertainties. It is only through
the methodology to be proposed below that an optimal `model trajectory' is
produced.


\subsubsection{Observable Parameters}

Given a complete model, as just defined, any observation can be predicted, 
as, for instance:
\begin{itemize}

\item
The reading of the time of a physical clock (on board of a satellite) can 
be predicted as \,$ z_i = \int_0^{\tau_i} d\tau \ 
\sqrt{g_{\alpha\beta} \, (d\tau^\alpha/d\tau) \, (d\tau^\beta/d\tau)} + f(\tau_i) $\, 
along the trajectory;

\item
The signals sent by the satellites may be observed by other satellites 
(this measurement being subject to experimental uncertainties). 
The time of arrival of the signals can be predicted by tracing the zero length 
geodesic going from the emission point to the reception trajectory (see
the methods proposed in appendix~\ref{sec: Arrival Time Data-2}).

\item
The satellites may have accelerometers, gradiometers, gyroscopes, etc.
The observations can also be computed using the given metric and the given 
trajectories.

\end{itemize}

The methodology here proposed in based in the assumption that 
\emph{any observation made by a satellite (the time of the physical clock,
the time of arrival of a received signal, the satellite acceleration, etc.) 
is broadcasted.} 
The goal of the paper is to propose a methodology that can allow any observer
to use all the broadcasted observations to build a complete model that 
is as good as possible. 
The model should predict values of the observations that are close to the
actual observations (within experimental uncertainties) and it should also
have some simple properties (for instance, the metric and the trajectories 
should have some degree of smoothness).
In principle, any observer could run his (inverse) modelization in real 
(proper) time.
As the accessible information will differ for the different observers,
the models will also differ. It is only for the part of the space-time 
that belongs to the common past of two observers that the models may be
arbitrarily similar (although not necessarily identical).

Note that the assumed smoothness of the metric and of the trajectories
will allow not only to obtain a model of space-time in the past of an observer,
but also in his future, although the accuracy of his prediction will
rapidly degrade with increasing prediction time (the methodology proposed
spontaneously evaluates uncertainties).

In some applications, the observer may not need to make a personal 
modelization, but just use a simple extrapolation of the information
about space-time broadcasted by a central observer in charge of all 
the computations%
\footnote{%
We call this central observer ``Houston''.}.
%
%


\subsection{First Constraint on the Metric (Zero Diagonal)}

It can be shown (see the lecture by J.M.\ Pozo in this school) that in the 
`light-coordinates' \,$ \{ \tau^\alpha \} $\, being used, the contravariant 
components of the metric must are have zeros on the diagonal,
\begin{equation}
\{ g^{\alpha\beta} \} \ = \ 
\begin{pmatrix}
0 & g^{12} & g^{13} & g^{14} \\
g^{12} & 0 & g^{23} & g^{24} \\
g^{13} & g^{23} & 0 & g^{34} \\
g^{14} & g^{24} & g^{34} & 0 \\
\end{pmatrix} \quad ,
\end{equation}
so the basic unknowns of the problem are the six quantities
\,$ \{ g^{12} , g^{13} , g^{14} , g^{23} , g^{24} , g^{34} \} $\,.
This constraint is imposed exactly, by just expressing all the relations
of the theory in terms of these six quantities%
\footnote{%
In the same lecture by J.M.\ Pozo, it is demonstrated that the metric above 
is Lorentzian if and only if the following condition is satisfied: defining 
\,$ A = \sqrt{g^{12} \, g^{34}} $\,, 
\,$ B = \sqrt{g^{13} \, g^{24}} $\,, and
\,$ C = \sqrt{g^{14} \, g^{23}} $\,, one must have
\,$ A + B > C $\,,
\,$ B + C > A $\,, and
\,$ C + A > B $\,. This constraint is not yet introduced, as it may be
strongly simplified when using the logarithmic metric.}.
%
%
The covariant components \,$ g_{\alpha\beta} $\, are defined, as usual, 
by the condition 
\,$ g_{\alpha\gamma} \, g^{\gamma\beta} = \delta_\alpha^\beta $\,.
The diagonal components of \,$ g_{\alpha\beta} $\, are \emph{not} zero.

\bigskip


\subsection{Second Constraint on the Metric (Smoothness)}

Let \,$ \bfg_\prior $\, be some simple initial estimation of the space-time
metric field. For instance, we could take for \,$ \bfg_\prior $\, the metric
of a flat space-time, the Schwarzschild metric of a point mass with the
Earth's mass, or a realistic estimation of the actual space-time metric 
around the Earth.

We wish that our final estimation of the metric, \,$ \bfg $\,, is
close to the initial estimation. More precisely, letting \,$ \bfC_\bfg $\, 
be a suitably chosen covariance operator,
we are going to impose that the least-squares norm%
\footnote{%
The criterion in equation~\protect\ref{eq: el-ese-del-g}, that is based on 
a difference of (contravariant) metrics, is only provisional. In a more 
advanced state of the theory, we should introduce the logarithm of the metric, 
and base the minimization criterion on the difference of logarithmic metrics.}
%
%
\begin{equation}
\Vert \, \bfg - \bfg_\prior \, \Vert_{\bfC_\bfg}^2 \ \equiv \
\langle \ \bfC_\bfg^{-1} \, ( \, \bfg - \bfg_\prior \, ) \ , \ 
( \, \bfg - \bfg_\prior \, ) \ \rangle
\label{eq: el-ese-del-g}
\end{equation}
is small.

The covariance operator, to be discussed later, shall be a 'smoothing operator'
this implying, from one side, that at every point of space-time the final
metric is close to the initial metric, and, from another side, that the
difference of the two metrics is smooth. As the initial metric shall be smooth,
this imposes that the final metric is also smooth. In particular, the final metric
will be defined `continously', in spite of the fact that we only `sample' it along
the space-time trajectories of the satellites and of the light signals.

This kind of smoothing, could perhaps be replaced by a criterion imposing that
the Riemann tensor should be as `small' as possible. The two possibilities must
be explored.


\subsection{Einstein Equation (Stress-Energy Data)}

The notations use in this text for the connection \,$ \Gamma^\alpha{}_{\beta\gamma} $\,,
the Riemann \,$ R^\alpha{}_{\beta\gamma\delta} $\,, and the Ricci \,$ R_{\alpha\beta} $\,
associated to a metric \,$ g_{\alpha\beta} $\,, are as follows:
\begin{equation}
\begin{split}
\Gamma^\alpha{}_{\beta\gamma} \ & = \ \tfrac{1}{2} \, g^{\alpha\sigma} \,
( \, \partial_\beta  \, g_{\gamma\sigma} + \partial_\gamma \, g_{\beta\sigma} 
- \partial_\sigma  \, g_{\beta\gamma} \, ) \\[3pt]
R^\alpha{}_{\beta\gamma\delta} \ & = \ 
\partial_\gamma  \, \Gamma^\alpha{}_{\delta\beta} 
- \partial_\delta  \, \Gamma^\alpha{}_{\gamma\beta} 
+ \Gamma^\alpha{}_{\mu\gamma} \, \Gamma^\mu{}_{\delta\beta}  
- \Gamma^\alpha{}_{\mu\delta} \, \Gamma^\mu{}_{\gamma\beta} \\[3pt]
R_{\alpha\beta} \ & = \ R^\gamma{}_{\alpha\gamma\beta} \quad . \\
\end{split}
\label{eq: bas-ei-eq}
\end{equation}
The Einstein tensor is then
\begin{equation}
E_{\alpha\beta} \ = \ R_{\alpha\beta} - \tfrac{1}{2} \, g_{\alpha\beta} \, R \quad ,
\label{eq: bas-ei-eq-0}
\end{equation}
where \,$ R = g^{\alpha\beta} \, R_{\alpha\beta} $\,.

The Einstein equation states that, at every point of the space-time, 
the Einstein tensor \,$ E_{\alpha\beta} $\, (associated to the metric) 
is proportional to the stress-energy tensor \,$ t_{\alpha\beta} $\, 
describing the matter at this space-time point:
\begin{equation}
E_{\alpha\beta} \ = \ \chi \, t_{\alpha\beta} \quad,
\end{equation}
where the proportionality constant is \,$ \chi = 8 \pi G/c^4 $\,. 
For instance, in vacuum, \,$ t_{\alpha\beta} = 0 $\,, and, therefore,
\,$ E_{\alpha\beta} = 0 $\,. When solving the Einstein equation for 
\,$ t_{\alpha\beta} $\,,
\begin{equation}
t_{\alpha\beta} \ = \ \frac{1}{\chi} \, E_{\alpha\beta}
\label{eq: stress-energy tensor}
\end{equation}
we obtain (when replacing \,$ E_{\alpha\beta} $\, by the 
expressions~\ref{eq: bas-ei-eq-0}--\ref{eq: bas-ei-eq}) the application
\begin{equation}
\bfg \quad \mapsto \quad \bft_\computed \ = \ \bft(\bfg) \quad ,
\label{eq: defino-t}
\end{equation}
associating to any metric field \,$ \bfg $\, the corresponding stress-energy
field \,$ \bft $\,.

Let \,$ \bft_\obs $\, be our estimation of the stress-energy of the space-time. 
It could, for instance, be zero, if we take for the space-time the model of 
vacuum. More realistically, we may take a simple model of the rarefied gas
that constitutes the high atmosphere. We wish that the space-time metric
\,$ \bfg $\, is such that the associated stress-energy \,$ \bft(\bfg) $\, is
close to \,$ \bft_\obs $\,.

More precisely, we are going to impose that the \,$ \bft(\bfg) - \bft_\obs $\,
is small in the sense of a least-squares norm
\begin{equation}
\Vert \, \bft(\bfg) - \bft_\obs \, \Vert_{\bfC_\bft}^2
\ \equiv \ \langle \ \bfC_\bft^{-1} \, ( \, \bft(\bfg) - \bft_\obs \, ) \ , \ 
( \, \bft(\bfg) - \bft_\obs \, ) \ \rangle
\quad ,
\label{eq: el-ese-del-E}
\end{equation}
where \,$ \bfC_\bft $\, is a covariance operator to be discussed later.
The notation \,$ \langle \, \cdot \, , \, \cdot \, \rangle $\, stands for
a duality product.

We shall later need the tangent linear application, \,$ \bfT $\,, to the 
application \,$ \bft(\bfg) $\,. By definition,
\begin{equation}
\bft( \, \bfg + \delta \bfg \, ) \ = \ \bft(\bfg) + \bfT \, \delta \bfg + \dots
\end{equation}
As demonstrated in appendix~\ref{sec: Perturbation of Einstein Tensor}
(see equations~\ref{eq: jmp-01}--\ref{eq: jmp-02}),
this linear tangent application is the (linear) application that
to every \,$ \delta g_{\alpha\beta} $\, associates the 
\,$ \delta t_{\alpha\beta} $\, given by
\begin{equation}
\delta t_{\alpha\beta} \ = \ \frac{1}{\chi} \, \big( \ 
A_{\alpha\beta}{}^{\mu\nu,\rho\sigma} \, \nabla_{\!(\rho}\nabla_{\!\sigma)} \, \delta g_{\mu\nu} 
+ B_{\alpha\beta}{}^{\mu\nu} \, \delta g_{\mu\nu} \ \big) \quad ,
\label{eq: Eva-T-k-1}
\end{equation}
where
\begin{equation}
\renewcommand\arraystretch{1.5}
\begin{array}{l@{{}={}}l}
A_{\alpha\beta}{}^{\mu\nu,\rho\sigma} & 
2 \, g^{(\mu |(\sigma}_{\phantom{)}} \, \delta^{\rho)}_{(\alpha } , \delta^{|\nu)}_{\beta)} 
- \frac{1}{2} \, g^{\mu\nu} \, \delta^\rho_{(\alpha } \, \delta^\sigma _{\beta)}
- \frac{1}{2} \, g^{\rho\sigma} \, \delta^\mu_{(\alpha } \, \delta^\nu_{\beta)}
+ \frac{1}{2} \, g^{\mu\nu} \, g^{\rho\sigma} \, g_{\alpha\beta}
- \frac{1}{2} \, g^{\mu (\rho} \, g^{\sigma )\nu} \, g_{\alpha\beta} \\
B_{\alpha\beta}{}^{\mu\nu} & \frac{1}{2} \,  (R^\mu {}_{(\alpha \beta)}{}^\nu 
+ R^{(\mu }{}_{(\alpha } \, \delta^{\nu)}_{\beta)}
+ R^{\mu\nu} \, g_{\alpha\beta} - R \, \delta^\mu_{(\alpha } \, \delta^\nu_{\beta)}) \quad .
\end{array}
\label{eq: Eva-T-k-2}
\end{equation}


\subsection{Proper Time Data}
\label{sec: Proper Time Data}

Assume that there is a physical clock (i.e., a perhaps accurate, but
certainly imperfect clock) on board of some of the satellites. 
At some instant along the trajectory of such a satelite, a reading 
of the physical clock is made (i.e., the proper time is measured), 
this giving some value \,$ z_{\rm obs} $\, with some associated uncertainty.

The theoretical prediction of the observation is made via the
integration of the space-time length element along trajectory.
So, given a model of the metric 
\,$ \bfg = \{ g_{\alpha\beta}(\tau^1,\tau^2,\tau^3,\tau^4)\} $\,, 
a model of the trajectory,
\,$ \bflambda = \{ \tau^\alpha(\lambda)\} $\,, and a model of the 
clock drift, \,$ \bff = \{ f(\tau) \} $\,, the prediction of the clock 
reading is
\begin{equation}
z(\tau) \ = \ \underbrace{\int_0^\tau \!\! d\lambda}_\bflambda \ \sqrt{g_{\alpha\beta} \, 
u^\alpha \, u^\beta}  + f(\tau) \quad ,  
\label{eq: one-single-t}
\end{equation}
where the integral is performed along the trajectory
where the vector \,$ u^\alpha $\, is defined (along the trajectory) as
\begin{equation}
u^\alpha \ = \ \frac{d\tau^\alpha}{d\lambda} \quad ,
\end{equation}
and where \,$ \lambda $\, is a parameter along the trajectory.

Expression~\ref{eq: one-single-t} is written for the evaluation of one single
time, while we shall typically many times evaluated along the trajectory,
\,$ \bfz = \{ z_1 , z_2 , \dots \} $\,. 
The application defined by expression~\ref{eq: one-single-t}, but considered for
all times, shall be written as
\begin{equation}
\{\bfg,\bflambda,\bff\} \ \mapsto \ \bfz_\text{computed} \ = \ 
\bfz(\bfg,\bflambda,\bff) \quad .
\end{equation}

Let \,$ \bfz_\text{obs} $\, the set of observed values, with experimental
uncertainties represented by a covariance matrix \,$ \bfC_\bfz $\,.
We wish that our final model \,$ \{ \bfg,\bflambda,\bff \} $\, is such that
the (least-squares) norm
\begin{equation}
\Vert \,  \bfz(\bfg,\bflambda,\bff) - \bfz_\text{obs} \, \Vert_{\bfC_\bfz}^2 \ \equiv \ 
\langle \ \bfC_\bfz^{-1} \, ( \, \bfz(\bfg,\bflambda,\bff) - \bfz_\text{obs} \, ) \ , \ 
( \, \bfz(\bfg,\bflambda,\bff) - \bfz_\text{obs} \, ) \ \rangle
\label{eq: misfit-one}
\end{equation}
is small.

For later use, we shall need the three (partial) linear tangent applications 
to the application so defined. They are defined through the series development
\begin{equation}
\bfz(\bfg+\delta\bfg,\bflambda+\delta\bflambda,\bff + \delta \bff) \ = \ 
\bfz(\bfg,\bflambda,\bff) + \bfZ_\bfg \, \delta\bfg + \bfZ_\bflambda \, \delta\bflambda 
+ \bfZ_\bff \, \delta \bff + \dots \quad ,
\end{equation}
where, for short, we write \,$ \bfZ_\bfg $\,, \,$ \bfZ_\bflambda $\,, and
\,$ \bfZ_\bff $\,, instead of
\,$ \bfZ_\bfg(\bfg,\bflambda,\bff) $\,, \,$ \bfZ_\bflambda(\bfg,\bflambda,\bff) $\,,
and \,$ \bfZ_\bff(\bfg,\bflambda,\bff) $\,.
One asily sees that \,$ \bfZ_\bfg $\, is the (linear) operator that
to any metric perturbation 
\,$ g_{\alpha\beta}(\tau^1,\tau^2,\tau^3,\tau^4) \mapsto 
g_{\alpha\beta}(\tau^1,\tau^2,\tau^3,\tau^4) + 
\delta g_{\alpha\beta}(\tau^1,\tau^2,\tau^3,\tau^4) $\,, 
associates, at each measure point, the time perturbation
\begin{equation}
\delta z \ = \ \frac{1}{2} \int d\lambda \ 
\frac{\delta g_{\alpha\beta} \, u^\alpha \, u^\beta}
{\sqrt{g_{\mu\nu} \, u^\mu \, u^\nu}} \quad .
\end{equation}
$ \bfZ_\bflambda $\, is the (linear) operator that
to any trajectory perturbation 
\,$ \tau^\alpha(\lambda) \mapsto 
\tau^\alpha(\lambda) + \delta \tau^\alpha(\lambda) $\,, 
associates, at each measure point, the time perturbation
\begin{equation}
\delta z \ = \ \int d\lambda \ 
\frac{g_{\alpha\beta} \, u^\alpha \, \delta u^\beta}
{\sqrt{g_{\mu\nu} \, u^\mu \, u^\nu}} \quad ,
\end{equation}
where
\begin{equation}
\delta u^\alpha \ = \ \frac{d \, \delta\tau^\alpha}{d\lambda} \quad .
\end{equation}
Finally, \,$ \bfZ_\bff $\, is the (linear) operator that to any perturbation
\,$ \delta f(\tau) $\, of the clock drift function associates
\begin{equation}
\delta z \ = \ \delta f \quad .
\end{equation}


\subsection{Arrival Time Data} 
\label{sec: Arrival Time Data-1} 

At some instant \,$ \tau_e $\, along its trajectory, a satellite 
emits a time signal, that is received by another satellite at 
(proper) time \,$ \sigma $\,.

Given a model of the metric 
\,$ \bfg = \{ g_{\alpha\beta}(\tau^1,\tau^2,\tau^3,\tau^4)\} $\,, 
a model of the trajectory that emits the signal,
\,$ \bftau_e = \{ \tau_e^\alpha(\tau)\} $\,, 
a model of the emission time \,$ \tau_e $\, along this trajectory, 
a model of the trajectory that receives the signal, 
\,$ \bftau_r = \{ \tau_r^\alpha(\tau)\} $\,, 
and a model of the clock drift of the receiving satellite,
\,$ \bff_r = \{ f_r(\tau) \} $\,, we can predict the reception
time by tracing the zero-length geodesic that connects the emission point 
to the receiving trajectory. We shall write this theoretical prediction as
\begin{equation}
\sigma_\text{computed} \ = \ \sigma(\bfg,\bftau_e,\tau_e,\bftau_r,\bff_r) \quad ,
\end{equation}
In the real situation, the
metric is only known approximately, and the computed value of the arrival
time will not be identical to the time
actually observed time, say \,$ \sigma_\obs $\,.

Roughly speaking, our goal is going to be to determine the space-time
metric that minimizes the differences between calculated and observed 
arrival times.

Our data, therefore, consists on a set of values
\begin{equation}
\{ \, \sigma_\obs^i \, \} \quad,
\end{equation}
assumed to be subjected to some observational uncertainties. 
Letting \,$ \bfC_\sigma $\, denote the covariance operator describing experimental
uncertainties, we wish the (least-squares) norm
\begin{equation}
\Vert \, \bfsigma(\bfg,\bftau_e,\tau_e,\bftau_r,\bff_r) - \bfsigma_\obs \, 
\Vert_{\bfC_\bfsigma}^2 \ \equiv \
\langle \ \bfC_\bfsigma^{-1} \, 
( \, \bfsigma(\bfg,\bftau_e,\tau_e,\bftau_r,\bff_r) - \bfsigma_\obs \, ) \ , \ 
( \, \bfsigma(\bfg,\bftau_e,\tau_e,\bftau_r,\bff_r) - \bfsigma_\obs \, ) \ \rangle
\label{eq: data-misfit-r}
\end{equation}
to be small.

Below, we shall need the (partial) tangent linear operators to the operator
\,$ \bfsigma $\,, defined as follows,
\begin{equation}
\begin{split}
& \bfsigma( \, \bfg+\delta\bfg \, , \, \bftau_e+\delta\bftau_e \, , \, 
\tau_e+\delta\tau_e \, , \, \bftau_r+\delta\bftau_r \, , \, 
\bff_r + \delta \bff_r \, ) \ = \ \\[5 pt]
& \bfsigma(\bfg,\bftau_e,\tau_e,\bftau_r,\bff_r) 
+ \bfSigma_{\bfg} \, \delta \bfg
+ \bfSigma_{\bftau_e} \, \delta \bftau_e
+ \bfSigma_{\tau_e} \, \delta \tau_e 
+ \bfSigma_{\bftau_r} \, \delta \bftau_r
+ \bfSigma_{\bff_r} \, \delta \bff_r
+ \dots \quad . \\
\end{split}
\end{equation}
Let us evaluate them.

When the metric is perturbed from 
\,$ g_{\alpha\beta} $\, to \,$ g_{\alpha\beta} + \delta g_{\alpha\beta} $\,,
the computed arrival times are perturbed from \,$ \sigma $\, to
\,$ \sigma + \delta\sigma $\,, where (see equation~\ref{eq: klimes-226} 
in appendix~\ref{sec: Arrival Time Data-2})
\begin{equation}
\delta \sigma \ = \ - \frac{1/2}{g_{\mu\nu} \, u^\mu \, \ell^\nu} \ 
\int_{\lambda(\bfg)} d\lambda \ \ell^\alpha \, \ell^\beta \, \delta g_{\alpha\beta}
\quad , 
\label{eq: lin-phi-337}
\end{equation}
where \,$ u^\alpha $\, is the tangent vector to the trajectory of the receiver,
\,$ u^\alpha = d\tau^\alpha/d\tau $\,, and \,$ \ell^\alpha $\, is the tangent vector
to the trajectory of the light ray, \,$ \ell^\alpha = d\tau^\alpha/d\lambda $\,
(where \,$ \lambda $\, is an affine parameter along the ray).
Therefore, \,$ \bfSigma_\bfg $\, is the linear operator that to any metric
perturbation \,$ \delta\bfg $\, associates the \,$ \delta\sigma $\, 
in equation~\ref{eq: lin-phi-337}.

To evaluate the operator \,$ \bfSigma_{\bftau_e} $\, we have to solve the
following problem: which is the first order perturbation \,$ \delta\sigma $\, 
to the arrival time when the trajectory of the emitter is perturbed from
\,$ \tau_e^\alpha(\tau) $\, to \,$ \tau_e^\alpha(\tau) + \delta \tau_e^\alpha(\tau)$\,? 
($ \to $ computation being performed).

To evaluate the operator \,$ \bfSigma_{\tau_e} $\, we have to solve the
following problem: which is the first order perturbation \,$ \delta\sigma $\, 
to the arrival time when the (proper) time the emitter is perturbed from
\,$ \tau_e $\, to \,$ \tau_e + \delta \tau_e $\,? 
($ \to $ computation being performed).

To evaluate the operator \,$ \bfSigma_{\bftau_r} $\, we have to solve the
following problem: which is the first order perturbation \,$ \delta\sigma $\, 
to the arrival time when the trajectory of the receiver is perturbed from
\,$ \tau_r^\alpha(\tau) $\, to \,$ \tau_r^\alpha(\tau) + \delta \tau_r^\alpha(\tau)$\,? 
($ \to $ computation being performed).

The operator \,$ \bfSigma_\bff $\, is the linear operator that to any
perturbation \,$ \delta f_r(\tau) $\, of the receiver clock drift functions
associates (check the notations)
\begin{equation}
\delta\sigma \ = \ \delta f_r(\sigma) \quad . 
\end{equation}


\subsection{Accelerometer Data}
\label{sec: Accelerometer Data}

We have to explore here the case where each `satellite' has an accelerometer.
The acceleration along a trajectory is
\begin{equation}
a^\alpha 
\ = \ u^\beta \, \frac{\partial u^\alpha}{\partial x^\beta} + 
\Gamma^\alpha{}_{\beta\gamma} \, u^\beta \, u^\gamma
\ = \ \frac{d u^\alpha}{d \tau} + 
\Gamma^\alpha{}_{\beta\gamma} \, u^\beta \, u^\gamma \quad ,
\label{eq: acceleration-Gamma}
\end{equation}
where \,$ \tau $\, is the proper time along the trajectory. 

The easiest way to `measure' the acceleration on-board would be, of course, 
to force the satellite (or its clock) to be in free-fall (i.e., to
follow a geodesic of the spacetime metric). Then, one would have \,$ a^\alpha = 0 $\,.
Let us keep considering here the general case where the acceleration may be
nonzero (because, for instance, by residual drag by the high atmosphere), 
but it is measured.

The measure of the acceleration provides information on the connection,
i.e., in fact, on the gradients of the metric.

Given a model \,$ \bfg $\, of the metric field and a model 
\,$ \bflambda $\, of the trajectory, equation~\ref{eq: acceleration-Gamma}
allows to compute the acceleration at all the space-time
points when it is measured. We write
\begin{equation}
\{\bfg,\bflambda\} \quad \mapsto \quad \bfa_\computed \ = \ \bfa(\bfg,\bflambda)
\label{eq: defino-a}
\end{equation}
the application so defined. We wish that the computed accelerations, 
\,$ \bfa(\bfg) $\,, are close to the observed ones, say \,$ \bfa_\obs $\,. 
More precisely, we wish the (least-squares) norm
\begin{equation}
\Vert \, \bfa(\bfg,\bflambda) - \bfa_\obs \, \Vert_{\bfC_\bfa}^2 \ \equiv \
\langle \ \bfC_\bfa^{-1} \, ( \, \bfa(\bfg,\bflambda) - \bfa_\obs \, ) \ , \ 
( \, \bfa(\bfg,\bflambda) - \bfa_\obs \, ) \ \rangle
\label{eq: data-misfit-a}
\end{equation}
to be small, where \,$ \bfC_\bfa $\, is a covariance operator describing the
experimental uncertainties in the measured acceleration values.

We introduce the tangent linear operators
\begin{equation}
\bfa( \, \bfg+\delta\bfg \, , \, \bflambda+\delta\bflambda \, ) \ = \ 
\bfa(\bfg,\bflambda) + \bfA_\bfg \, \delta \bfg + \bfA_\lambda \, \delta \bflambda
+ \dots
\label{eq: perturbing-acceleration}
\end{equation}

It follows from equation~\ref{eq: acceleration-Gamma} that a perturbation of
the metric \,$ g_{\alpha\beta} \mapsto g_{\alpha\beta} + \delta g_{\alpha\beta} $\,,
produces a perturbation of the computed acceleration given by
\,$ \delta a^\alpha = \delta \Gamma^\alpha{}_{\beta\gamma} \, 
u^\beta \, u^\gamma $\,.
The expression for \,$ \delta \Gamma^\alpha{}_{\beta\gamma} $\, is in 
appendix~\ref{sec: Perturbation of Einstein Tensor}
(see equation~\ref{eq: delta-Gamma}, page~\pageref{eq: delta-Gamma}),
\,$ \delta\Gamma^\alpha{}_{\beta\gamma} = 
g^{\alpha\sigma} \, \delta\Gamma_{\sigma\beta\gamma} $\,,
with
\,$ \delta\Gamma_{\alpha\beta\gamma} = 
\tfrac{1}{2} \, (\nabla_{\!\gamma} \delta g_{\alpha\beta} 
+ \nabla_{\!\beta} \delta g_{\alpha\gamma}-\nabla_{\!\alpha} \delta g_{\beta\gamma}) $\,.
This gives
\begin{equation}
\delta a^\alpha \ = \ \tfrac{1}{2} \, g^{\alpha\sigma} \, 
\big( \nabla_{\!\gamma} \, \delta g_{\sigma\beta} 
+ \nabla_{\!\beta} \, \delta g_{\sigma\gamma} 
- \nabla_{\!\sigma} \, \delta g_{\beta\gamma} \big) \, u^\beta \, u^\gamma 
\quad . 
\label{eq: delta-a-alpha}
\end{equation}
The linear operator so defined was denoted \,$ \bfA_\bfg $\, 
in equation~\ref{eq: perturbing-acceleration}. To any metric
field perturbation \,$ \delta g_{\alpha\beta} $\, this operator associates,
at every point of a space-time trajectory where the acceleration was
measured, the values \,$ \delta a^\alpha $\, just written.

We leave to the reader the characterization of the operator \,$ \bfA_\bflambda $\,.


\subsection{Gyroscope Data}
\label{sec: Gyroscope Data}

A gyroscope is described by its \emph{spin vector} (or angular momentum vector) 
\,$ s^{\alpha} $\,, a four-vector that is orthogonal to the four-velocity 
\,$ u^\alpha $\, of the rotating particle: 
\,$ g_{\alpha\beta} \, u^\alpha \, s^\beta $\,.

Assume that the gyroscope follows a trajectory \,$ x^\alpha = x^\alpha(\tau) $\,,
whose velocity is \,$ u^\alpha = dx^\alpha/d\tau $\, and whose acceleration 
\,$ a^\alpha $\, is that expressed in equation~\ref{eq: acceleration-Gamma}.
Then, the evolution of the spin vector along the trajectory is described%
\footnote{%
For details on the relativistic treatment of a spinning test particle, see
Papatetrou (1951), Weinberg (1972), or Hern\'andez-Pastora et al. (2001).}
%
%
by the so-called Fermi-Walker transport:
\begin{equation}
\frac{Ds^\alpha}{d\tau} \ \equiv \ 
\frac{ds^\alpha}{d\tau} + \Gamma^\alpha{}_{\beta\gamma} \, u^\beta \, s^\gamma \ = \ 
s_\beta \, ( a^\beta \, u^\alpha - a^\alpha \, u^\beta ) \quad .
\end{equation}
Should the gyroscope be in free fall, \,$ a^\alpha = 0 $\,, and 
\,$ ds^\alpha/d\tau + 
\Gamma^\alpha{}_{\beta\gamma} \, u^\beta \, s^\gamma = 0 $\,, 
this meaning that the spin vector would be transported by parallelism.

In our case, the monitoring of the spin vector \,$ s^\alpha(\tau) $\, 
(besides the monitoring of the acceleration \,$ a^\alpha $\,) would provide the
values \,$ \Gamma^\alpha{}_{\beta\gamma} \, u^\beta \, s^\gamma $\,, an information
complementary to that provided by the monitoring of the acceleration (that provides
the values  \,$ \Gamma^\alpha{}_{\beta\gamma} \, u^\beta \, u^\gamma $\,). 

Consider that our data is 
\begin{equation}
\pi^\alpha \ = \ \frac{ds^\alpha}{d\tau} \quad .
\end{equation}
Then we have
\begin{equation}
\pi^\alpha  \ = \ 
s_\beta \, ( a^\beta \, u^\alpha - a^\alpha \, u^\beta ) 
- \Gamma^\alpha{}_{\beta\gamma} \, u^\beta \, s^\gamma
\quad .
\label{eq: spin-Gamma}
\end{equation}

Given the metric field model \,$ \bfg $\, and the trajectory model \,$ \bflambda $\,, 
this equation
allows to compute the vector \,$ \pi^\alpha $\, at all the space-time
points when it is measured. We write
\begin{equation}
\{\bfg,\bflambda\} \quad \mapsto \quad \bfpi_\computed \ = \ \bfpi(\bfg,\bflambda)
\label{eq: defino-s}
\end{equation}
the application so defined. We wish that the computed values, 
\,$ \bfpi(\bfg) $\,, are close to the observed ones, say \,$ \bfpi_\obs $\,. 
More precisely, we wish the (least-squares) norm
\begin{equation}
\Vert \, \bfpi(\bfg,\bflambda) - \bfpi_\obs \, \Vert_{\bfC_\bfpi}^2 \ \equiv \
\langle \ \bfC_\bfpi^{-1} \, ( \, \bfpi(\bfg,\bflambda) - \bfpi_\obs \, ) \ , \ 
( \, \bfpi(\bfg,\bflambda) - \bfpi_\obs \, ) \ \rangle
\label{eq: data-misfit-s}
\end{equation}
to be small, where \,$ \bfC_\bfpi $\, is a covariance operator describing the
experimental uncertainties in the measured values.

Of course, one may not wish to measure the evolution of the spin vector to provide
information on the connection, but to `test' general relativity, as in the Gravity 
Probe B experiment. From the viewpoint of the present work, the detection of any 
inconsistency in the data would put relativity theory in jeopardy.

Let us introduce the linear tangent operators
\begin{equation}
\bfpi( \, \bfg +\delta\bfg \, , \, \bflambda + \delta\bfg \, ) \ = \ 
\bfpi(\bfg,\bflambda) + \bfPi_\bfg \, \delta\bfg 
+ \bfPi_\bflambda \, \delta\bflambda + \dots
\end{equation}

The application \,$ \bfg \mapsto \bfpi(\bfg) $\, 
is given in equation~\ref{eq: spin-Gamma}.
To compute the first order perturbation \,$ \bfpi \mapsto\bfpi + \delta\bfpi $\,
produced by a perturbation \,$ \bfg \mapsto \bfg + \delta\bfg $\,, we must, 
in this equation, make the replacements 
\,$ \Gamma^\alpha{}_{\beta\gamma} \mapsto \Gamma^\alpha{}_{\beta\gamma} + \delta \Gamma^\alpha{}_{\beta\gamma} $\,
and \,$ s^\alpha \mapsto s^\alpha + \delta s^\alpha $\,, with subsequent expression
of \,$ \delta \Gamma^\alpha{}_{\beta\gamma} $\, and \,$ \delta s^\alpha $\, in
terms of \,$ \delta g_{\alpha} $\,. 
We obtain
\begin{equation}
\delta \pi^\alpha \ = \ - \delta \Gamma^\alpha{}_{\beta\gamma} \, u^\beta \, s^\gamma
\quad .
\end{equation}
Using the expression for \,$ \delta \Gamma^\alpha{}_{\beta\gamma} $\, 
in appendix~\ref{sec: Perturbation of Einstein Tensor},
we are immediately left to an expression similar to~\ref{eq: delta-a-alpha}:
\begin{equation}
\delta \pi^\alpha \ = \ - \tfrac{1}{2} \, g^{\alpha\sigma} \, 
\big( \nabla_{\!\gamma} \, \delta g_{\sigma\beta} 
+ \nabla_{\!\beta} \, \delta g_{\sigma\gamma} 
- \nabla_{\!\sigma} \, \delta g_{\beta\gamma} \big) \, u^\beta \, \sigma^\gamma 
\quad . 
\label{eq: delta-pi-alpha}
\end{equation}
The operator that to any \,$ \delta g_{\alpha\beta} $\, associates the
\,$ \delta \pi^\alpha $\, given by this equation is the operator 
\,$ \bfPi_\bfg $\,, we were searching for.

We leave to the reader the evaluation of the operator \,$ \bfPi_\bflambda $ \,.


\subsection{Gradiometer Data}
\label{sec: Gradiometer Data}

To study the gravity field around the Earth, different satellite missions 
are on course or planned%
\footnote{%
The LAGEOS (LAser GEOdynamics Satellites) are passive spherical bodies covered
with retroreflectors. Note that Ciufolini and Pavlis (2004) have recently
been able to confirm the Lense-Thirring effect using LAGEOS data.
The CHAMP (CHAllenging Minisatellite Payload) satellite is equipped with a 
precise orbit determination and an accelerometer.
The GRACE (GRAvity recovery and Climate Experiment) consists in two satellites
with precise orbit determination, accelerometers and measure of their mutual
distance with an accuracy of a few microns.
The GOCE (Gravity Field and Steady-State Ocean Circulation Explorer) satellite 
has recently been launched. It consists in a three axis \emph{gradiometer}:
six accelerometers in a so-called diamond configuration. The observables are 
the differences of the accelerations.}.
%
%
Of particular importance are the \emph{gradiometers} with which modern gravimetric
satellites are equipped. In the GOCE satellite, there are three 
perpendicular ``gradiometer arms'', each arm consisting in two masses 
(50 cm apart) that are submitted to electrostatic forces to keep each of 
them at the center of a cage. These forces are monitored, thus providing
the accelerations. The basic data are the half-sum and the difference of 
these accelerations (for each of the three gradiometer arms). 

The half-sum of the accelerations gives what a simple accelerometer would give.
The difference corresponds to the ``tidal forces'' in the region where the
satellite operates. 

\begin{figure}[htbp] 
   \centering
   \includegraphics[width=25mm]{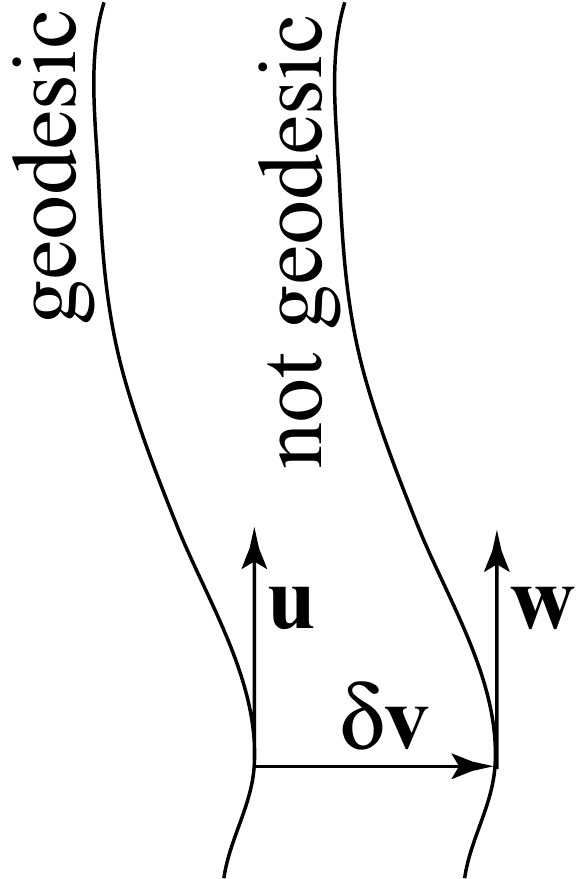} 
   \caption{For the incorporation of gradiometry data, we consider a geodesic 
space-time trajectory, and the trajectory defined by transporting a small vector
along the geodesic (see text for details).}
   \label{fig: Gradiometry}
\end{figure}

A simple model for the gradiometry data is as follows. A mass follows some
space-time line that, to simplify the discussion, is assumed to be a geodesic
(i.e., the mass is assumed to be in free-fall, but taking into account its 
possible acceleration would be simple). 
(We leave to the reader the writing of the general formulas for the case when the initial trajectory is not a geodesic.)
This geodesic is represented at the 
left in figure~\ref{fig: Gradiometry}. Let \,$ u^\alpha $\, be the unit vector
tangent to this geodesic trajectory. Consider, at some initial point along the
geodesic, a ``small'' space-time vector \,$ \delta v^\alpha $\, that, 
to fix ideas, may be assumed to be a space-like vector. By parallel transport 
of \,$ \delta v^\alpha $\, along the geodesic one defines a second trajectory, 
that is not necessarily a geodesic (the line at the right in 
figure~\ref{fig: Gradiometry}). Let us denote \,$ w^\alpha $\, the tangent 
vector to this trajectory, and \,$ \delta a^\alpha $\, the acceleration along it.
Note that, as the trajectory is close to being geodesic, the acceleration 
\,$ \delta a^\alpha $\, is small (and would vanish if \,$ \delta v^\alpha = 0 $\,).


Notice that one has
\begin{equation}
g_{\mu\nu} \, u^\mu \, u^\nu \ = \ 1 \qquad ; \qquad 
g_{\mu\nu} \, w^\mu \, \delta a^\nu \ = \ 0 \quad ,
\end{equation}
and that (i) the tangent vector \,$ w^\alpha $\, is obtained, all along
the trajectory, by parallel transport of \,$ u^\alpha $\, along 
\,$ \delta v^\alpha $\,, and (ii) at this level of approximation, 
the proper time along the second trajectory is identical to the proper time
along the first trajectory.

A mass can be forced to follow this line, and the forces required to do this
can be monitored, this giving a measurement of the acceleration 
\,$ \delta a^\alpha $\, of the mass.

We do not need to exactly evaluate the theoretical relation expressing 
\,$ \delta a^\alpha $\,, the approximation that is first order in 
\,$ \delta v^\alpha $\, will be sufficient (because \,$ \delta v^\alpha $\, 
is small). As demonstrated by Pozo et al.~(2005), one has 
\,$ \delta a^\alpha = 
R^\alpha{}_{\mu\nu\rho} \, u^\mu \, u^\rho \, \delta v^\nu + \dots $\,,
where the remaining terms are at least second order in \,$ \delta v^\alpha $\,.
Then, with a sufficient approximation, we use below the expression
\begin{equation}
\delta a^\alpha \ = \  
R^\alpha{}_{\mu\nu\rho} \, u^\mu \, u^\rho \, \delta v^\nu \quad .
\label{eq: Constraints-Riemann}
\end{equation}
As the three vectors \,$ a^\alpha $\,, \,$ u^\alpha $\,, and 
\,$ \delta v^\alpha $\, are known, we have a direct information on the 
components of the Riemann tensor.

A typical gradiometer contains three arms (in three perpendicular directions
in space). This means that we have three different vectors \,$ \delta v^\alpha $\,
with which to apply equation~\ref{eq: Constraints-Riemann}. The vector 
\,$ u^\alpha $\, is unique (fixed by the trajectory of the satellite). 
Should one have different satellites at approximately the same space-time point,
with significantly different trajectories, one would have extra constraints
on the Riemann tensor (at the given space-time point).

In order to simplify the notations in later sections of the paper, we drop the 
\,$ \delta $\, for the vector \,$ \delta v^\alpha $\,, and we write 
\,$ \omega^\alpha $\, instead of \,$ \delta a^\alpha $\,. Then, 
equation~\ref{eq: Constraints-Riemann} becomes
\begin{equation}
\omega^\alpha \ = \ 
R^\alpha{}_{\mu\nu\rho} \, u^\mu \, u^\rho \, v^\nu \quad . 
\label{eq: Constraints-Riemann-2}
\end{equation}

Given a model metric field \,$ \bfg $\, and a model trajectory \,$ \bflambda $\,, 
the theoretical values of the tidal acceleration are detoted \,$ \bfomega_\computed $\,, 
and we write
\begin{equation}
\{\bfg,\bflambda\} \quad \mapsto \quad \bfomega_\computed \ = \ \bfomega(\bfg,\bflambda) \quad ,
\label{eq: defino-omega}
\end{equation}
where \,$ \omega^\alpha_\computed = 
R^\alpha{}_{\mu\nu\rho}(\bfg) \, u^\mu \, u^\rho \, v^\nu $\,.
The gradiometer provides the `observed acceleration' \,$ \bfomega_\obs $\,,
with observational uncertainties represented by a covariance operator
\,$ {\bf C}_\bfomega $\,. We wish that the tidal accelerations, 
\,$ \bfomega(\bfg,\bflambda) $\,, are close to the observed ones, \,$ \bfomega_\obs $\,. 
More precisely, we wish the (least-squares) norm
\begin{equation}
\Vert \, \bfomega(\bfg,\bflambda) - \bfomega_\obs \, \Vert_{\bfC_\bfomega}^2 \ \equiv \
\langle \ \bfC_\bfomega^{-1} \, ( \, \bfomega(\bfg,\bflambda) - \bfomega_\obs \, ) \ , \ 
( \, \bfomega(\bfg,\bflambda) - \bfomega_\obs \, ) \ \rangle
\label{eq: data-misfit-alpha}
\end{equation}
to be small.

We introduce the linear tangent operators
\begin{equation}
\bfomega( \, \bfg + \delta\bfg \, , \, \bflambda + \delta \bflambda \, ) \ = \ 
\bfomega(\bfg,\bflambda) + \bfOmega_\bfg \, \delta \bfg 
+ \bfOmega_\bflambda \, \delta \bflambda + \dots
\end{equation}

In view of equation~\ref{eq: Constraints-Riemann-2}, a perturbation of the metric
field will produce the perturbation
\begin{equation}
\delta\omega^\alpha \ = \ 
\delta R^\alpha{}_{\beta\gamma\delta} \, u^\beta \, u^\delta \, v^\gamma 
\label{eq: bf-Omega-1}
\end{equation}
of the tidal acceleration, where \,$ \delta R^\alpha{}_{\mu\nu\rho} $\, 
is the first order perturbation to the Riemann tensor. 
This perturbation is obtained as a by product in our computation of the perturbation
of the Einstein tensor in appendix~\ref{sec: Perturbation of Einstein Tensor}
(see equation~\ref{eq: First-order-Riemann}, page~\pageref{eq: First-order-Riemann}).
The result is
\begin{equation}
\delta R^\alpha{}_{\beta\gamma\delta} \ = \ 
2\nabla_{\![\gamma}\Omega^\alpha{}_{\delta]\beta} \quad ,
\end{equation}
where
\begin{equation}
\Omega^\alpha{}_{\beta\gamma} \ = \ g^{\alpha\sigma}\Omega_{\sigma\beta\gamma}
\qquad\text{with}\qquad 
\Omega_{\alpha\beta\gamma} \ = \
\tfrac{1}{2} \, (\nabla_{\!\gamma} \delta g_{\alpha\beta} 
+ \nabla_{\!\beta} \delta g_{\alpha\gamma}-\nabla_{\!\alpha} \delta g_{\beta\gamma}) 
\quad .
\label{eq: bf-Omega-3}
\end{equation}
This characterizes the operator \,$ \bfOmega_\bfg $\,.

We leave to the reader the characterization of the operator \,$ \bfOmega_\bflambda $\,.


\subsection{Total Misfit}
\label{sec: Total Misfit}

It is clear that we need to optimize for all the components of a complete model,
and these include, in particular, the space-time metric and the space-time
trajectories. In the context of this school, we choose to present a simplified
version of the theory, were only the space-time metric is optimized. The students
should be able to complete the expression of the total misfit, as an exercise.

Using standard arguments from least-squares theory (see Tarantola [2005]),
we shall define here the 
`best metric field' as the field \,$ \bfg $\, that minimizes the
sum of all the misfit terms introduced above (equations~\ref{eq: el-ese-del-E},
\ref{eq: el-ese-del-g}, \ref{eq: misfit-one}, \ref{eq: data-misfit-r},
\ref{eq: data-misfit-a}, \ref{eq: data-misfit-s}, and~\ref{eq: data-misfit-alpha}).
The total misfit function, that we denote \,$ S(\bfg) $\,, is, therefore, 
given by
\begin{equation}
\boxed{ \qquad
\begin{split}
2 \, S(\bfg) \ = \ 
& \Vert \, \bfg - \bfg_\prior \, \Vert_{\bfC_\bfg}^2 +
\Vert \,  \bfz(\bfg) - \bfone \, \Vert_{C_\bfz}^2 +
\Vert \, \bft(\bfg) - \bft_\obs \, \Vert_{\bfC_\bft}^2 +
\Vert \, \bfsigma(\bfg) - \bfsigma_\obs \, \Vert_{\bfC_\bfsigma}^2 \\[3 pt]
+ &
\Vert \, \bfa(\bfg) - \bfa_\obs \, \Vert_{\bfC_\bfa}^2 + 
\Vert \, \bfpi(\bfg) - \bfpi_\obs \, \Vert_{\bfC_\bfpi}^2 +
\Vert \, \bfomega(\bfg) - \bfomega_\obs \, \Vert_{\bfC_\bfomega}^2 \quad , \\
\end{split}
\quad } 
\label{eq: total misfit}
\end{equation}
i.e.,
\begin{equation}
\begin{split}
2 \, S(\bfg) \ = \ \ 
& \langle \ \bfC_\bfg^{-1} \, ( \, \bfg - \bfg_\prior \, ) \ , \ 
( \, \bfg - \bfg_\prior \, ) \ \rangle \\[3 pt]
+ & \langle \ \bfC_\bfz^{-1} \, ( \, \bfz(\bfg) - \bfone \, ) \ , \ 
( \, \bfz(\bfg) - \bfone \, ) \ \rangle \\[3 pt]
+ & \langle \ \bfC_\bft^{-1} \, ( \, \bft(\bfg) - \bft_\obs \, ) \ , \ 
( \, \bft(\bfg) - \bft_\obs \, ) \ \rangle \\[3 pt]
+ & \langle \ \bfC_\bfsigma^{-1} \, ( \, \bfsigma(\bfg) - \bfsigma_\obs \, ) \ , \ 
( \, \bfsigma(\bfg) - \bfsigma_\obs \, ) \ \rangle \\[3 pt]
+ & \langle \ \bfC_\bfa^{-1} \, ( \, \bfa(\bfg) - \bfa_\obs \, ) \ , \ ( \, \bfa(\bfg) - \bfa_\obs \, ) \ \rangle \\[3 pt]
+ & \langle \ \bfC_\bfpi^{-1} \, ( \, \bfpi(\bfg) - \bfpi_\obs \, ) \ , \ ( \, \bfpi(\bfg) - \bfpi_\obs \, ) \ \rangle \\[3 pt]
+ & \langle \ \bfC_\bfomega^{-1} \, ( \, \bfomega(\bfg) - \bfomega_\obs \, ) \ , \ 
( \, \bfomega(\bfg) - \bfomega_\obs \, ) \ \rangle \quad . \\
\end{split}
\label{eq: total misfit-2}
\end{equation}

Sometimes, in least-squares theory it is allowed for these different terms to
have different `weights', by multiplying them by some ad-hoc numerical factors.
This is not necessary if all the covariance operators are chosen properly.
In any case, adding some extra numerical factors is a trivial task that we do
not contemplate here.

Although in this paper we limit our scope to providing the simplest method 
that could be used to actually find the metric field \,$ \bfg $\, that
minimizes the misfit function, it is interesting to know that the function
\,$ S(\bfg) $\, carries a more fundamental information. In fact, as
shown, for instance, in Tarantola (2005), the expression
\begin{equation}
\varphi(\bfg) \ = \ k \, \exp( \, - \, S(\bfg) \, ) 
\label{eq: total misfit proba}
\end{equation}
defines a probability density (infinite-dimensional) that represents the 
information we have on the actual metric field, i.e., in fact, the respective 
`likelihoods' of all possible metric fields.

\bigskip


\section{Optimization}

\def\toto{\medskip}

\toto

We are going to present here a very plain optimization algorithm, based on the
Newton's method. This is not a valid candidate for a practical algorithm. 
Let us see why.

\toto

The Newton algorithm, as proposed here, produces an ab-initio calculation:
one takes all the observations, the a priori information on the complete model
and produces the a posteriori information on the complete model. All the data
are treated together. Of course, any practical algorithm should rather use the
basic idea of a \emph{Kalman filter}: data are integrated into the computation
as they are available (see appendix~\ref{sec: Kalman Filter} for a short
description of the linear Kalman filter). 

\toto

In itself, this in an important topic for our future research: 
Kalman filter computations are made in real time, but what is `real time'
in the context of the relativistic physics used here? It turns out that
`real time' is replaced in this context by `proper observer time' and that
the data integrated by an observer in his Kalman filter are the data arriving
to him at this moment, i.e., the data belonging to the `surface' of his 
past light cone. We leave this advanced topic out from the teachings of this
school.

\toto

Therefore, we proceed with the analysis of a simple steepest-descent algorithm.


\toto

\toto

\subsection{Iterative Algorithm}
\label{sec: Iterative Algorithm}

\toto

Once the misfit function \,$ S(\bfg) $\, as been defined 
(equation~\ref{eq: total misfit-2}), and the associated probability distribution
\,$ f(\bfg) $\, has been introduced (equation~\ref{eq: total misfit proba}),
the ideal (although totally impractical) approach for extracting all the 
information on \,$ \bfg $\, brought by the data of our problem would be to 
sample the probability distribution%
\footnote{%
Sampling an infinite-dimensional probability distribution is not possible,
but we could define a (dense enough) grid in the space-time where the
values of \,$ \bfg $\, are considered, this discretization rendering the
probability distribution finite-dimensional.}
%
%
\,$ f(\bfg) $\,. Examples of the sampling of a probability distribution 
in the context of inverse problems can be found in Tarantola (2005).

In the present problem, where the initial metric shall not be too far from
the actual metric, the nonlinearities of the problem are going to be weak.
This implies that the probability distribution
\,$ f(\bfg) $\, is monomodal, i.e., the misfit function \,$ S(\bfg) $\,
has a unique minimum (in the region of interest of the parameter space).
Therefore, the general sampling techniques can here be replaced by the
much more efficient optimization techniques. The basic question becomes:
{\sl for which metric field \,$ \bfg $\, the misfit function \,$ S(\bfg) $\,
attains its minimum?}

\toto

This problem can be solved using gradient-based techniques. These techniques
are quite sophisticated, and require careful adaptation to the problem at
hand if they have to work with acceptable efficiency. As we do not wish to
develop this topic in this paper, we just choose here to explore the 
simple steepest descent algorithm, while we explore the more complete
Newton algorithm in appendix~\ref{sec: Newton Algorithm}.

\toto

To run a steepest descent optimization algorithm, there is only one
evaluation that must be done extremely accurately, the evaluation of the
direction of steepest ascent. For this one may use the formulas developed in 
Tarantola (2005). One obtains the following direction of steepest ascent,
\begin{equation}
\boxed{ \qquad
\begin{split}
\bfgamma_k \ = \ (\bfg_k - \bfg_\prior) 
& + (\bfZ_k \, \bfC_\bfg)^t \, \bfC_\bfz^{-1} \, (\bfz(\bfg_k) - \bfone) \\[3 pt]
& + (\bfT_k \, \bfC_\bfg)^t \, \bfC_\bft^{-1} \, (\bft(\bfg_k) - \bft_\obs) \\[3 pt]
& + (\bfSigma_k \, \bfC_\bfg)^t \, 
                \bfC_\bfsigma^{-1} \, (\bfsigma(\bfg_k) - \bfsigma_\obs) \\[3 pt]
& + (\bfA_k \, \bfC_\bfg)^t \, \bfC_\bfa^{-1} \, (\bfa(\bfg_k) - \bfa_\obs) \\[3 pt]
& + (\bfPi_k \, \bfC_\bfg)^t \, \bfC_\bfpi^{-1} \, (\bfpi(\bfg_k) - \bfpi_\obs) \\[3 pt]
& + (\bfOmega_k \, \bfC_\bfg)^t \, 
                     \bfC_\bfomega^{-1} \, (\bfomega(\bfg_k) - \bfomega_\obs) \quad , \\
\end{split}
\quad } 
\label{eq: gradient vector-new}
\end{equation}
where the linear operators \,$ \bfZ_k $\,, \,$ \bfT_k $\,, \,$ \bfSigma_k $\,, 
\,$ \bfA_k $\,, \,$ \bfPi_k $\,, and \,$ \bfOmega_k $\, have all
been introduced above.
The meaning of \,$ ( \cdot )^t $\, in the above should be obvious.
We say \emph{transpose} operators, better than \emph{dual}
operators, because the difference between the two notions matters inside the
theory of least-squares%
\footnote{%
In fact, the dual operators (denoted with a `star') ere respectively
\,$ \bfZ_k^\ast = \bfC_\bfg \, \bfZ_k^t \, \bfC_\bfz^{-1} $\,,
\,$ \bfT_k^\ast = \bfC_\bfg \, \bfT_k^t \, \bfC_\bft^{-1} $\,,
\,$ \bfSigma_k^\ast = \bfC_\bfg \, \bfSigma_k^t \, \bfC_\bfsigma^{-1} $\,, 
\,$ \bfA_k^\ast = \bfC_\bfg \, \bfA_k^t \, \bfC_\bfa^{-1} $\,, 
\,$ \bfPi_k^\ast = \bfC_\bfg \, \bfPi_k^t \, \bfC_\bfpi^{-1} $\,, and 
\,$ \bfOmega_k^\ast = \bfC_\bfg \, \bfOmega_k^t \, \bfC_\bfomega^{-1}$\,.
See Tarantola (2005) for details.}.
%
%

The Newton algorithm presented in appendix~\ref{sec: Newton Algorithm},
is typically not used as such. One rather uses a `preconditioned steepest
descent algorithm',
\begin{equation}
\boxed{\qquad
\bfg_{k+1} \ = \ \bfg_k - \bfP_k \, \bfgamma_k \quad , \quad } 
\end{equation}
where \,$ \bfP_k $\, is an ad-hoc positive definite operator, suitably chosen
to produce a convergence of the algorithm as rapid as possible.
Should one choose to use for \,$ \bfP_k $\, the Hessian of the misfit function,
one would obtain the Newton algorithm.

Alternatively, one may choose to use a `relaxation algorithm', where succesive
`jumps' are performed along the different directions defined by the different
terms in equation~\ref{eq: gradient vector-new}.

One should keep in mind that to obtain a proper estimation of the posterior uncertainties
in the metric, one needs the evaluation of the inverse of the Hessian operator
(see appendix~\ref{sec: Newton Algorithm}).


\subsection{Transpose of a Linear Application}
\label{sec: Transpose of a Linear Application}

Let \,$ \bfL $\, be a linear operator mapping a linear space
\,$ \SpaceA $\, into a linear space \,$ \SpaceB $\,. Let \,$ \SpaceA^\ast $\,
and \,$ \SpaceB^\ast $\, the respective dual spaces, and
\,$ \langle \, \cdot \, , \, \cdot \, \rangle_\SpaceA $\, and
\,$ \langle \, \cdot \, , \, \cdot \, \rangle_\SpaceB $\, the respective
duality products. The transpose of \,$ \bfL $\, is the linear operator mapping
\,$ \SpaceB^\ast $\, into \,$ \SpaceA^\ast $\, such that for any
\,$ \bfa \in \SpaceA $\, and any \,$ \bfb^\ast \in \SpaceB^\ast $\,,
\begin{equation}
\langle \, \bfb^\ast \, , \, \bfL \, \bfa \, \rangle_\SpaceB \ = \ 
\langle \, \bfL^t \, \bfb^\ast \, , \, \bfa \, \rangle_\SpaceA \quad .
\end{equation}
For a good text on functional analysis, in particular on the 
transpose and adjoint of a linear operator, see Taylor and Lay (1980).
Some of the results in the following sections are provided without
demonstration: to check the proposed results, the reader should 
become familiar with the concepts proposed in that book.
Let us only mention here two elementary results. 
The transpose of the linear operator defined through
\begin{equation}
y^{ij\dots}{}_{k\ell\dots} \ = \ 
A^{ij\dots\mu\nu\dots}{}_{k\ell\dots\alpha\beta\dots} \, x^{\alpha\beta\dots}{}_{\mu\nu\dots\dots}
\end{equation}
is the linear operator defined through
\begin{equation}
x_{\alpha\beta\dots}{}^{\mu\nu\dots\dots} \ = \ 
A^{ij\dots\mu\nu\dots}{}_{k\ell\dots\alpha\beta\dots} \, 
y_{ij\dots}{}^{k\ell\dots}
\end{equation}
The transpose of the linear operator defined through
\begin{equation}
y^{\alpha\beta\dots}{}_{\gamma\mu\nu\dots\dots} \ = \ 
\nabla_\gamma \, x^{\alpha\beta\dots}{}_{\mu\nu\dots\dots}
\end{equation}
is the linear operator defined through
\begin{equation}
x_{\alpha\beta\dots}{}^{\mu\nu\dots\dots} \ = \ 
- \nabla_\gamma \, y_{\alpha\beta\dots}{}^{\gamma\mu\nu\dots\dots} 
\end{equation}
There are typically some boundary conditions to be attached to a differential
operator, what implies for the transpose operator a set of `dual' boundary
conditions, but we shall not enter into these `details' in this preliminary
version of the theory.


\subsubsection{Einstein Equation}
\label{sec: Appendix Einstein Equation}

Considering the two duality products
\begin{equation}
\begin{split}
\langle \, \delta\hat{\bft} \, , \, \delta \bft \, \rangle \ & = \ 
\int d\tau^1\int d\tau^2\int d\tau^3\int d\tau^4
 \ \delta\hat{t}^{\alpha\beta}(\tau^1,\tau^2,\tau^3,\tau^4) \, 
 \delta t_{\alpha\beta}(\tau^1,\tau^2,\tau^3,\tau^4) \\
\langle \, \delta\hat{\bfg} \, , \, \delta \bfg \, \rangle \ & = \ 
\int d\tau^1\int d\tau^2\int d\tau^3\int d\tau^4
 \ \delta\hat{g}^{\alpha\beta}(\tau^1,\tau^2,\tau^3,\tau^4) \, \delta g_{\alpha\beta}(\tau^1,\tau^2,\tau^3,\tau^4) \quad , \\
\end{split}
\end{equation}
the transpose operator \,$ \bfT^t $\, is defined by the condition that for any
\,$ \delta\hat{\bft} $\, and for any \,$ \delta\bfg $\,, one must have
\begin{equation}
\langle \, \bfT^t \delta\hat{\bft} \, , \, \delta\bfg \, \rangle \ = \ 
\langle \, \delta\hat{\bft} \, , \, \bfT \, \delta\bfg \, \rangle \quad .  
\end{equation}
Using equation~\ref{eq: Eva-T-k-1} and the expression of the duality products,
this can be written
\begin{equation}
\begin{split}
& \int d\tau^1\int d\tau^2\int d\tau^3\int d\tau^4 \
\left[ \bfT^t \, \delta\hat{\bft} \right]^{\alpha\beta} \ \delta g_{\alpha\beta} \ = \ \\
& \frac{1}{\chi} \int d\tau^1\int d\tau^2\int d\tau^3\int d\tau^4 \ 
\delta\hat{t}^{\alpha\beta} \ \big( \ 
A_{\alpha\beta}{}^{\mu\nu,\rho\sigma} \, \nabla_{\!(\rho}\nabla_{\!\sigma)} \, \delta g_{\mu\nu} 
+ B_{\alpha\beta}{}^{\mu\nu} \, \delta g_{\mu\nu} \ \big) \quad . 
\end{split}
\label{eq: from-met-toC55}
\end{equation}
To compute the direction of steepest descent we need to evaluate the term
\,$ \bfC_\bfg \, \bfT^t \, \delta \hat{\bft} $\,, i.e., we need to evaluate
\begin{equation}
\begin{split}
& \left[ \bfC_\bfg \, \bfT^t \, \delta \hat{\bft} \right]_{\gamma\delta}
(\sigma^1,\sigma^2,\sigma^3,\sigma^4) \\
& = \ 
\int d\tau^1\int d\tau^2\int d\tau^3\int d\tau^4 \ 
\left[ \bfT^t \, \delta\hat{\bft} \right]^{\alpha\beta}
(\tau^1,\tau^2,\tau^3,\tau^4) \ 
C_{\alpha\beta\gamma\delta}(\tau^1,\tau^2,\tau^3,\tau^4,
\sigma^1,\sigma^2,\sigma^3,\sigma^4) \quad . \\
\end{split}
\end{equation}
This can be evaluated by just replacing 
\,$ \delta g_{\alpha\beta}(\tau^1,\tau^2,\tau^3,\tau^4) $\,
by \,$ C_{\alpha\beta\gamma\delta}(\tau^1,\tau^2,\tau^3,\tau^4,
\sigma^1,\sigma^2,\sigma^3,\sigma^4) $\, in equation~\ref{eq: from-met-toC55}. 
One gets
\begin{equation}
\begin{split}
& \left[ \bfC_\bfg \, \bfT^t \, \delta \hat{\bft} \right]_{\gamma\delta}
(\sigma^1,\sigma^2,\sigma^3,\sigma^4) \ = \ \\[5 pt]
& \frac{1}{\chi} \int d\tau^1\int d\tau^2\int d\tau^3\int d\tau^4 \ 
\delta\hat{t}^{\alpha\beta}(\tau^1,\tau^2,\tau^3,\tau^4) \\
& \Big( \ 
A_{\alpha\beta}{}^{\mu\nu,\rho\sigma}(\tau^1,\tau^2,\tau^3,\tau^4) \ \nabla_{\!(\rho}\nabla_{\!\sigma)} \, C_{\mu\nu\gamma\delta}
(\tau^1,\tau^2,\tau^3,\tau^4,\sigma^1,\sigma^2,\sigma^3,\sigma^4) \\[5 pt]
& + B_{\alpha\beta}{}^{\mu\nu}(\tau^1,\tau^2,\tau^3,\tau^4) \ 
C_{\mu\nu\gamma\delta}(\tau^1,\tau^2,\tau^3,\tau^4,\sigma^1,\sigma^2,\sigma^3,\sigma^4) \ \Big) \quad . 
\end{split}
\label{eq: from-met-toC555}
\end{equation}


\subsubsection{Arrival Time Data}
\label{sec: Arrival Time Data}

For each satellite trajectory, to any space-time metric field perturbation 
\,$ \delta g_{\alpha\beta}(\tau^1,\tau^2,\tau^3,\tau^4) $\,, 
expression~\ref{eq: lin-phi-337} associates the scalar \,$ \delta\sigma(\tau) $\,
defined along the trajectory.
Let us here characterize the transpose, \,$ \bfSigma^t $\,, of this operator.
Considering the two duality products
\begin{equation}
\begin{split}
\langle \, \delta\hat{\bfsigma} \, , \, \delta \bfsigma \, \rangle \ & = \ 
\int d\tau \ \delta\hat{\sigma}(\tau) \, \delta \sigma(\tau) \\
\langle \, \delta\hat{\bfg} \, , \, \delta \bfg \, \rangle \ & = \ 
\int d\tau^1\int d\tau^2\int d\tau^3\int d\tau^4
 \ \delta\hat{g}^{\alpha\beta}(\tau^1,\tau^2,\tau^3,\tau^4) \, \delta g_{\alpha\beta}(\tau^1,\tau^2,\tau^3,\tau^4) \quad , \\
\end{split}
\end{equation}
the transpose operator is defined by the condition that for any
\,$ \delta\hat{\bfsigma} $\, and for any \,$ \delta\bfg $\,, one must have
\begin{equation}
\langle \, \bfSigma^t \delta\hat{\bfsigma} \, , \, \delta\bfg \, \rangle \ = \ 
\langle \, \delta\hat{\bfsigma} \, , \, \bfSigma \, \delta\bfg \, \rangle \quad .  
\end{equation}
Using equation~\ref{eq: lin-phi-337} and the expression of the duality products,
this can be written
\begin{equation}
\begin{split}
& 
\int d\tau^1\int d\tau^2\int d\tau^3\int d\tau^4 \ 
\left[ \bfSigma^t \, \delta\hat{\bfsigma} \right]^{\alpha\beta}
(\tau^1,\tau^2,\tau^3,\tau^4) \ \delta g_{\alpha\beta}(\tau^1,\tau^2,\tau^3,\tau^4) \\
& = \ \int d\tau \ \delta\hat{\sigma}(\tau) \ 
\left( - \frac{1/2}{g_{\mu\nu}(\tau^\kappa(\tau)) \, u^\mu(\tau) \, 
\ell^\nu(\tau) }  \ \int_{\lambda(\bfg,\tau)} d\lambda \ \ell^\alpha(\lambda) \, 
\ell^\beta(\lambda) \, \delta g_{\alpha\beta}(\tau^\kappa(\lambda))\right) \quad . \\
\end{split}
\label{eq: from-met-toC}
\end{equation}
To compute the direction of steepest descent we need to evaluate the term
\,$ \bfC_\bfg \, \bfSigma^t \, \delta \hat{\bfsigma} $\,, i.e., we need to evaluate
\begin{equation}
\begin{split}
& \left[ \bfC_\bfg \, \bfSigma^t \, \delta \hat{\bfsigma} \right]_{\gamma\delta}
(\sigma^1,\sigma^2,\sigma^3,\sigma^4) \\
& = \ 
\int d\tau^1\int d\tau^2\int d\tau^3\int d\tau^4 \ 
\left[ \bfSigma^t \, \delta\hat{\bfsigma} \right]^{\alpha\beta}
(\tau^1,\tau^2,\tau^3,\tau^4) \ 
C_{\alpha\beta\gamma\delta}(\tau^1,\tau^2,\tau^3,\tau^4,
\sigma^1,\sigma^2,\sigma^3,\sigma^4) \quad . \\
\end{split}
\end{equation}
This can be evaluated by just replacing 
\,$ \delta g_{\alpha\beta}(\tau^1,\tau^2,\tau^3,\tau^4) $\,
by \,$ C_{\alpha\beta\gamma\delta}(\tau^1,\tau^2,\tau^3,\tau^4,
\sigma^1,\sigma^2,\sigma^3,\sigma^4) $\, in equation~\ref{eq: from-met-toC}. 
One gets
\begin{equation}
\begin{split}
& \left[ \bfC_\bfg \, \bfSigma^t \, \delta \hat{\bfsigma} \right]_{\gamma\delta}
(\sigma^1,\sigma^2,\sigma^3,\sigma^4) \ = \ \int d\tau \ \delta\hat{\sigma}(\tau) \ 
\left( - \frac{1/2}{g_{\mu\nu}(\tau^\kappa(\tau)) \, u^\mu(\tau) \, 
\ell^\nu(\tau) }  \right. \\
& \left . \int_{\lambda(\bfg,\tau)} d\lambda \ \ell^\alpha(\lambda) \, 
\ell^\beta(\lambda) \, C_{\alpha\beta\gamma}(\tau^1(\lambda),\tau^2(\lambda),\tau^3(\lambda),\tau^4(\lambda),
\sigma^1,\sigma^2,\sigma^3,\sigma^4)\right) \quad . \\
\end{split}
\end{equation}
As the covariance function \,$ C_{\alpha\beta\gamma\delta}(\tau^1,\tau^2,\tau^3,\tau^4,
\sigma^1,\sigma^2,\sigma^3,\sigma^4) $\, shall be a smooth function, 
we see that this equation `spreads' the arrival time residuals into the region of the
space-time that is around each satellite trajectory.


\subsubsection{Accelerometer Data}

It follows from equation~\ref{eq: delta-a-alpha}
that the transpose operator 
\,$ \bfA_\bfg^t $\, is the linear operator that to any \,$ \delta \hat{a}_\alpha $\,,
defined at the points where the acceleration was measured, associates, at the
same points, the values
\begin{equation}
\boxed{ \qquad
\delta\hat{g}^{\alpha\beta} \ = \ - \tfrac{1}{2} \,
\big( g^{\alpha\nu} \, u^\beta \, u^\mu + g^{\beta\nu} \, u^\alpha \, u^\mu 
- g^{\mu\nu} \, u^\alpha \, u^\beta \big) \ \nabla_\mu \, \delta\hat{\pi}_\nu
\quad . \quad } 
\end{equation}
This operator appears in equations~\ref{eq: gradient vector 2} 
and~\ref{eq: final linear system}.


\subsubsection{Gyroscope Data}

The operator, \,$ \bfPi_\bfg^t $\,, transpose of the operator \,$ \bfPi_\bfg $\, 
characterized in equation~\ref{eq: delta-pi-alpha} is 
\begin{equation}
\boxed{ \quad
\delta\hat{g}^{\alpha\beta} \ = \ - \tfrac{1}{4} \,
\big( g^{\alpha\nu} \, (u^\beta \, \sigma^\mu + u^\mu \, \sigma^\beta) 
+ g^{\beta\nu} \, (u^\alpha \, \sigma^\mu + u^\mu \, \sigma^\alpha) 
- g^{\mu\nu} \, (u^\alpha \, \sigma^\beta + u^\beta \, \sigma^\alpha) \big) \ \nabla_\mu \, \delta\hat{\pi}_\nu
\ \ . \ \ } 
\end{equation}
It appears in equations~\ref{eq: gradient vector 2} 
and~\ref{eq: final linear system}.


\subsubsection{Gradiometer Data}

The operator \,$ \bfOmega_\bfg $\, was characterized in 
equations~\ref{eq: bf-Omega-1}--\ref{eq: bf-Omega-3}.
The transpose operator, \,$ \bfOmega_\bfg^t $\,, associates to any 
\,$ \delta\hat{\omega}_\alpha $\, the \,$ \delta\hat{g}^{\alpha\beta} $\, given by
\begin{equation}
\begin{split}
\delta\hat{g}^{\alpha\beta} \ = \
& v^\mu \, u^\nu \, ( 
  u^\alpha \, \nabla_{\mu\nu} \delta\hat{\omega}^\beta 
+ u^\beta \, \nabla_{\mu\nu} \delta\hat{\omega}^\alpha ) 
+ (u^\alpha \, v^\beta + u^\beta \, v^\alpha) \, u^\mu \, \nabla_{\mu\nu} 
\delta\hat{\omega}^\nu \\
- & u^\mu \, u^\nu \, ( v^\alpha \, \nabla_{\mu\nu} 
\delta\hat{\omega}^\beta + 
v^\beta \, \nabla_{\mu\nu} \delta\hat{\omega}^\alpha 
- 2 \, u^\alpha \, u^\beta \, v^\mu \, 
\nabla_{\mu\nu} \delta\hat{\omega}^\nu \quad .
\end{split}
\end{equation}
It appears in equations~\ref{eq: gradient vector 2} 
and~\ref{eq: final linear system}.


\section{Discussion and Conclusion}

We have been able to develop a consistent theory, fully relativistic, 
where the data brought by satellites emitting and receiving time signals 
is used to infer trajectories and the space-time metric. This constitutes
both, a kind of ultimate gravimeter and a positioning system. Any observer
with receiving capabilities shall know its own space-time trajectory 
``in real time''. These coordinates are not the usual `geographical'
coordinates plus a time, but are four times. The problem of attaching these
four time coordinates to any terrestrial system of coordinates is just an
attachment problem that should not interfere with the basic problem of
defining an accurate reference system, and of knowing space-time trajectories 
into this system.

For more generality, we have considered the possibility that the satellites
may have accelerometers, gradiometers, or gyroscopes. This is because the
positioning problem and the problem of estimating the gravity field (i.e.,
the space-time metric) are coupled. If fact, all modern gravimetry satellite
missions are coupled with GNSS satellites. Our theory applies, in particular,
to the GOCE satellite mission (orbiting gradiometers). It also applies
to the Gravity Probe B or the LISA%
\footnote{%
Laser Interferometer Space Antenna.}
%
%
experiments, that could be analyzed using the concepts presented here.

The optimization algorithm proposed (Newton algorithm) is by no means the 
more economical to be used in the present context, and considerable effort
is required to propose a practical algorithm, possibly using the `Kalman
filter' approach briefly mentioned in appendix~\ref{sec: Kalman Filter}.
We are quite confident in our prediction that, some day, all positioning
systems will be run using the basic principles exposed in this paper:
the ever-increasing accuracy of time measurements with eventually force
everyone to take relativity theory seriously ---at last.--- 


\section{Bibliography}

\def\bibi{\par\noindent\hangindent=20pt}

\bibi
Ashby, N., 2002, 
Relativity and the global positioning system, 
Physics Today, 55 (5), May 2002, pp.\ 41--47.

\bibi
Blagojevi\'c, M., J.\ Garecki, F.W.\ Hehl, and Yu.N.\ Obukhov, 2001, Real null coframes in general relativity and GPS type coordinates, arXiv:gr-qc/0110078v3.

\bibi
Ciarlet, P.G., 1982, Introduction \`a l'analyse num\'erique matricielle et \`a 
l'optimisation, Masson, Paris.

\bibi
\v{C}erven\'{y},~V., 2002,
Fermat's variational principle for anisotropic inhomogeneous media,
Stud.\ geophys.\ geod., vol.\ 46, pp.\ 567--588. 
({\tt http://sw3d.mff.cuni.cz}).

\bibi
Ciufolini, I., and Pavlis, E.C., 2004,
A confirmation of the general relativistic prediction of the
Lense-Thirring effect, Nature, Vol.\ 431, pp.\ 958--960.

\bibi
Coll, B., 1985, Coordenadas luz en relatividad, Proc.\ Spanish Relativity Meeting, ERE-85, Publ.\ Servei Publ.\ ETSEIB, Barcelona, p 29--39. English translation (Light coordinates in relativity) at {\tt http://

\vskip -0.4 mm\hskip 1 mm syrte.obspm.fr/{$\sim$}coll/Papers/CoordinateSystems/LightCoordinates.pdf}

\bibi
Coll, B., and Morales, J.A., 1992,
199 causal classes of space-time frames,
International Journal of Theoretical Physics, Vol.\ 31, No.\ 6,
pp.\ 1045--1062.

\bibi
Coll, B., 2000, 
Elements for a theory of relativistic coordinate systems, 
formal and Physical aspects, ERES 2000, Valladolid.

\bibi
Coll, B., 2001a, Elements for a Theory of Relativistic
Coordinate Systems; Formal and Physical Aspects, Proc. XXIII Spanish 
Relativity Meeting ERE2000, World Scientific.

\bibi
Coll, B., 2001b, 
Physical Relativistic Frames, JSR 2001,
ed.\ N.\ Capitaine, Pub.\ Observatoire de Paris, pp.\ 169--174.

\bibi
Coll, B., 2002, 
A principal positioning system for the Earth, JSR 2002, 
eds.\ N.\ Capitaine and M.\ Stavinschi, Pub.\ Observatoire de Paris, pp.\ 34--38.

\bibi
Coll, B.\ and Tarantola, A., 2003,
Galactic positioning system; physical relativistic coordinates for the Solar system
and its surroundings, eds.\ A.\ Finkelstein and N.\ Capitaine,
JSR 2003, Pub.\ St.\ Petersbourg Observatory, pp.\ 333-334.

\bibi
Gordon, W., 1923, Ann.\ Phys.\ (Leipzig), 72, p.\ 421.

\bibi
Grewal, M.S., Weill, L.R., and Andrews, A.P., 2001, Global positioning systems, 
inertial navigation, and integration, John Wiley \& Sons.

\bibi
Hern\'andez-Pastora, J.L., Mart\'\i n, J., and Ruiz, E., 2001,
On gyroscope precession, arXiv:gr-qc/0009062.

\bibi
Klime\v{s},~L., 2002,
Second-order and higher-order perturbations of travel time in isotropic and 
ani\-so\-tro\-pic media,
Stud.\ geophys.\ geod., vol.\ 46, pp.\ 213--248.
({\tt http://sw3d.mff.cuni.cz}).

\bibi
Papapetrou, A., 1951, 
Spinning test-particles in general relativity (I),
Proceedings of the Royal Society of London.\ A209, pp.\ 248--258

\bibi
Pham Mau Quan, 1957, Arch.\ Rat.\ Mech.\ and Anal., 1, p.\ 54.

\bibi
Rovelli, C., 2002, GPS observables in general relativity, Phys.\ Rev.\ D 65, 044017. Posted in the arXiv in 2001 as arXiv:gr-qc/0110003. 

\bibi
Tarantola, A., 2005, Inverse problem theory and methods for model parameter estimation,
SIAM.

\bibi
Taylor, A.E., and Lay, D.C., 1980,
Introduction to functional analysis, Wiley.

\bibi
Weinberg, S., 1972, Gravitation and Cosmology, Wiley.


\section{Appendixes}


\subsection{Perturbation of Einstein's Tensor}
\label{sec: Perturbation of Einstein Tensor}

Note that, as for any matrix \,$ \bfa $\,,
\,$ (\bfa + \delta \bfa)^{-1} = 
\bfa^{-1} - \bfa^{-1} \, \delta\bfa \, \bfa^{-1} + \cdots $\,,
when imposing to the metric the perturbation
\begin{equation}
g_{\alpha\beta} \ \mapsto \ g_{\alpha\beta} + \delta g_{\alpha\beta} \quad ,
\label{eq: per-met-1}
\end{equation}
the contravariant components have the perturbation
\begin{equation}
g^{\alpha\beta} \ \mapsto \ g^{\alpha\beta} - g^{\alpha\gamma} \, \delta g_{\gamma\delta} \, g^{\delta\beta} + \cdots \quad .
\label{eq: per-met-2}
\end{equation}

When introducing the perturbations~\ref{eq: per-met-1}--\ref{eq: per-met-2}
in the expressions~\ref{eq: bas-ei-eq}, one obtains, keeping only first
order terms in \,$ \delta g_{\alpha\beta} $\,, the perturbation \,$ \delta E_{\alpha\beta} $\, 
of the Einstein tensor.

We consider the perturbation of the metric and the perturbation of the 
connection. 
\begin{equation}
g_{\alpha\beta} \ \to \ g'_{\alpha\beta} \ = \ g_{\alpha\beta} + \delta g_{\alpha\beta} \qquad ; \qquad
\Gamma^\alpha{}_{\beta\gamma} \ \to \ 
\Gamma'^\alpha{}_{\beta\gamma} \ = \ \Gamma^\alpha{}_{\beta\gamma} + \delta\Gamma^\alpha{}_{\beta\gamma} \quad .
\end{equation}
In this appendix, and in order to make the expressions more compact, 
let us denote 
\begin{equation}
\delta g_{\alpha\beta} \ = \ h_{\alpha\beta} \qquad ; \qquad 
\delta\Gamma^\alpha{}_{\beta\gamma} \ = \ \Omega^\alpha{}_{\beta\gamma} \quad .
\end{equation}
We will use the unperturbed metric to raise and lower indices. For 
instance, we will write $h^\alpha{}_\beta \equiv g^{\alpha\gamma}h_{\gamma\beta}$\,.

By requiring that both, the unperturbed and the perturbed connection, to be 
metric, $\nabla g=\nabla'g'=0$, and symmetric, $\Omega^\alpha{}_{[\beta\gamma]}=0$, 
we get:
\begin{equation}
\left. \begin{array}{r}
\nabla'_{\!\gamma} g'_{\alpha\beta}= \nabla_{\!\gamma} h_{\alpha\beta}-\Omega^\delta{}_{\alpha\gamma}g'_{\delta\beta}
-\Omega^\delta{}_{\beta\gamma}g'_{\alpha\delta}=0 \\ 
\Omega^\alpha{}_{[\beta\gamma]}=0
\end{array} \right\}
\ \Rightarrow\ 
g'_{\alpha\delta}\Omega^\delta{}_{\beta\gamma}=
\frac{1}{2} \, (\nabla_{\!\gamma} h_{\alpha\beta}+\nabla_{\!\beta} h_{\alpha\gamma}-\nabla_{\!\alpha} h_{\beta\gamma}) \ .
\end{equation}
Then, $\Omega^\mu{}_{\beta\gamma}$ is obtained by contracting this 
expression with the inverse of the perturbed metric $g'^{\mu\alpha}$. But the first 
order is given by the contraction with the unperturbed one:
\begin{equation}
\Omega^\alpha{}_{\beta\gamma}=g^{\alpha\delta}\Omega_{\delta \beta\gamma}
\qquad\mbox{where}\qquad 
\Omega_{\alpha\beta\gamma}=
\frac{1}{2} \, (\nabla_{\!\gamma} h_{\alpha\beta}+\nabla_{\!\beta} h_{\alpha\gamma}-\nabla_{\!\alpha} h_{\beta\gamma}) \ .
\label{eq: delta-Gamma}
\end{equation}
The two Riemann tensors are related by
\begin{equation}
R'^\alpha{}_{\beta\gamma\delta}=R^\alpha{}_{\beta\gamma\delta}+
2\nabla_{\![\gamma}\Omega^\alpha{}_{\delta]\beta}+2\Omega^\alpha{}_{\mu[\gamma}\Omega^\mu{}_{\delta]\beta} \ .
\end{equation}
Thus, the first order perturbation of the Riemann is given by
\begin{equation}
\delta R^\alpha{}_{\beta\gamma\delta}=2\nabla_{\![\gamma}\Omega^\alpha{}_{\delta]\beta} \ ,
\label{eq: First-order-Riemann}
\end{equation}
from which we get the first order perturbation of the Ricci tensor:
\begin{equation}
\delta R_{\alpha\beta}=\delta R^\delta{}_{\alpha\delta\beta}=
2\nabla_{\![\delta}\Omega^\delta{}_{\beta]\alpha}=
\nabla_{\![\delta}\nabla_{\!\beta]}h^\delta{}_\alpha +
\frac{1}{2} \, (\nabla_{\!\delta}\nabla_{\!\alpha}h^\delta{}_\beta 
-\nabla_{\!\beta}\nabla_{\!\alpha}h^\delta{}_\delta)
-\frac{1}{2} \, (\nabla_{\!\delta}\nabla^\delta h_{\alpha\beta}
+\nabla_{\!\beta}\nabla^\delta h_{\alpha\delta}) \ .
\end{equation}
Splitting into symmetric and intisymmetric parts of the two covariant 
derivatives 
\begin{equation}
\renewcommand\arraystretch{1.5}
\begin{array}{r @{}l}
\delta R_{\alpha\beta} = {} &
\nabla_{\![\delta}\nabla_{\!\beta]}h^\delta{}_\alpha 
+ \frac{1}{2} \, \nabla_{\![\delta}\nabla_{\!\alpha]}h^\delta{}_\beta 
+ \frac{1}{2} \, \nabla_{\![\beta}\nabla_{\!\delta]} h^\delta{}_\alpha  \\
& {}+ \frac{1}{2} \, \nabla_{\!(\delta}\nabla_{\!\alpha)}h^\delta{}_\beta 
- \frac{1}{2} \, \nabla_{\!(\beta}\nabla_{\!\alpha)}h^\delta{}_\delta
- \frac{1}{2} \, \nabla_{\!\delta}\nabla^\delta h_{\alpha\beta}
+ \frac{1}{2} \, \nabla_{\!(\beta}\nabla_{\!\delta)} h^\delta{}_\alpha  \ .
\end{array}
\end{equation}
Substituting now the identity 
$2\nabla_{\![\delta}\nabla_{\!\alpha]}h^\delta{}_\beta 
=R^\delta{}_{\mu \delta \alpha} \, h^\mu{}_\beta -R^\mu{}_{\beta\delta \alpha} \, h^\delta{}_\mu $
and introducing the notation 
\begin{equation}
H_{\alpha\beta,\gamma\delta} \ \equiv \ \nabla_{\!(\gamma}\nabla_{\!\delta)}h_{\alpha\beta} \quad ,
\label{eq: H-as-h}
\end{equation}
\begin{equation}
2 \, \delta R_{\alpha\beta} = 
R_{\mu (\alpha}h^\mu{}_{\beta )}+R^\mu{}_{(\alpha\beta)\delta}h^\delta{}_\mu 
+H^\delta{}_{\beta ,\alpha\delta}+H^\delta{}_{\alpha,\beta\delta}-H^\delta{}_{\delta,\alpha\beta}
-H_{\alpha\beta,}{}^\delta{}_\delta \ .
\end{equation}

In order to obtain the perturbation of the Ricci scalar we also need the 
perturbation of the contravariant metric: 
\begin{equation}
g^{\alpha\gamma  }\, g_{\gamma\beta}=\delta^\alpha_\beta  \ \Rightarrow\ 
\delta g^{\alpha\beta}=-g^{\alpha\gamma} \, h_{\gamma\delta} \, g^{\delta\beta}=-h^{\alpha\beta} \ .
\end{equation}
Thus, the perturbation of the Ricci scalar is
\begin{equation}
\delta R=\delta g^{\alpha\beta} \, R_{\alpha\beta}+g^{\alpha\beta} \, \delta R_{\alpha\beta}=
-R^{\alpha\beta} \, h_{\alpha\beta}+H^{\alpha\beta}{}_{,\alpha\beta}-H^\alpha{}_{\alpha,}{}^\beta{}_\beta  \ .
\end{equation}
Finally we obtain the first order perturbation of the Einstein tensor,
\begin{equation}
\renewcommand\arraystretch{1.5}
\begin{array}{r @{}l}
\delta E_{\alpha\beta} ={} & \delta R_{\alpha\beta}-\frac{1}{2} \, (\delta R \, g_{\alpha\beta}
+ R \, \delta g_{\alpha\beta}) \\
={} & \frac{1}{2} \, (R^\delta{}_{(\alpha\beta)}{}^\mu  \, h_{\delta \mu}+R^\mu{}_{(\alpha} \, h_{\beta )\mu}
+R^{\gamma\delta} \, h_{\gamma\delta} \, g_{\alpha\beta} - R \, h_{\alpha\beta}) \\
& {}+\frac{1}{2} \, (H^\delta{}_{\beta ,\alpha\delta}+H^\delta{}_{\alpha,\beta\delta}
-H^\delta{}_{\delta,\alpha\beta}-H_{\alpha\beta,}{}^\delta{}_\delta
+H^\gamma {}_{\gamma ,}{}^\delta{}_\delta g_{\alpha\beta} - H^{\gamma\delta}{}_{,\gamma\delta} \, g_{\alpha\beta})
\end{array}
\end{equation}
This result can be rewriten as
\begin{equation}
\delta E_{\alpha\beta} \ = \ A_{\alpha\beta}{}^{\gamma\delta,\rho\sigma} \, H_{\gamma\delta,\rho\sigma} 
+ B_{\alpha\beta}{}^{\gamma\delta} \, h_{\gamma\delta}
\label{eq:195729}
\end{equation}
with
\begin{equation}
\renewcommand\arraystretch{1.5}
\begin{array}{l@{{}={}}l}
A_{\alpha\beta}{}^{\gamma\delta,\rho\sigma} & 
2 \, g^{(\gamma|(\sigma}_{\phantom{)}} \, \delta^{\rho )}_{(\alpha} , \delta^{|\delta)}_{\beta )} 
- \frac{1}{2} \, g^{\gamma\delta} \, \delta^\rho_{(\alpha} \, \delta^\sigma _{\beta )}
- \frac{1}{2} \, g^{\rho\sigma} \, \delta^\gamma _{(\alpha} \, \delta^\delta_{\beta )}
+ \frac{1}{2} \, g^{\gamma\delta} \, g^{\rho\sigma} \, g_{\alpha\beta}
- \frac{1}{2} \, g^{\gamma (\rho} \, g^{\sigma )\delta} \, g_{\alpha\beta} \\
B_{\alpha\beta}{}^{\gamma\delta} & \frac{1}{2} \,  (R^\gamma {}_{(\alpha\beta)}{}^\delta 
+ R^{(\gamma}{}_{(\alpha} \, \delta^{\delta)}_{\beta )}
+ R^{\gamma\delta} \, g_{\alpha\beta} - R \, \delta^\gamma _{(\alpha} \, \delta^\delta_{\beta )}) \ .
\end{array}
\label{eq: jmp-01}
\end{equation}
Note that using he definition~\ref{eq: H-as-h}, equation~\ref{eq:195729} 
can be written, explicitly,
\begin{equation}
\delta E_{\alpha\beta} = A_{\alpha\beta}{}^{\gamma\delta,\rho\sigma} \, \nabla_{\!(\rho}\nabla_{\!\sigma)} h_{\gamma\delta} 
+ B_{\alpha\beta}{}^{\gamma\delta} \, h_{\gamma\delta} \quad . 
\label{eq: jmp-02}
\end{equation}

Observe that, by construction, the two tensors \,$ {\bf A} $\, and 
\,$ {\bf B} $\, are symmetrics in each pair of indices,
\begin{equation}
A_{\alpha\beta}{}^{\gamma\delta,\mu\nu} = A_{(\alpha\beta)}{}^{(\gamma\delta),(\mu\nu)} \qquad\mbox{and}\qquad
B_{\alpha\beta}{}^{\gamma\delta} = B_{(\alpha\beta)}{}^{(\gamma\delta)} \ .
\end{equation}
In addition, it results that $A_{\alpha\beta}{}^{\gamma\delta,\mu\nu}$ is symmetric respect to the 
interchange of the two contravariant pairs:
\begin{equation}
A_{\alpha\beta}{}^{\gamma\delta,\mu\nu} = A_{\alpha\beta}{}^{\mu\nu,\gamma\delta} \ .
\end{equation}
This implies that not all the information in $H_{\alpha\beta,\gamma\delta}$ contrubutes to 
$\delta E_{\alpha\beta}$. In fact, we can express the term 
$A_{\alpha\beta}{}^{\gamma\delta,\mu\nu}H_{\alpha\beta,\gamma\delta}$ 
in an interesting form. Let us define $J_{\alpha\beta\gamma\delta}\equiv 2H_{[\delta|[\alpha,\beta]|\gamma]}$.
This tensor contains less information than $H_{[\gamma|[\alpha,\beta]|\delta]}$, and has the 
same symmetries as a Riemann:
\begin{equation}
J_{\alpha\beta\gamma\delta}=J_{[\alpha\beta][\gamma\delta]}=J_{\gamma\delta \alpha\beta} 
\qquad\mbox{and}\qquad
J_{\alpha[\beta\gamma\delta]}=0 \ .
\end{equation}
We can then take the traces of this tensor (obtaining a Ricci-like 
tensor and scalar): $J_{\alpha\beta}\equiv J^\gamma {}_{\alpha\gamma \beta}$ and $\beta\equiv J^\alpha{}_\alpha $. 
Then, it is easy to check that the contribution of $H_{\alpha\beta,\gamma\delta}$ is only the 
Einstein-like tensor of $J_{\alpha\beta\gamma\delta}$:
\begin{equation}
A_{\alpha\beta}{}^{\gamma\delta,\mu\nu}H_{\gamma\delta,\mu\nu} \ = \ 
J_{\alpha\beta}-\frac{1}{2} \,  J \, g_{\alpha\beta}
\end{equation}
In contrast, observe that $B_{\alpha\beta}{}^{\gamma\delta}$ contains all the information of 
the Riemann tensor.


\subsection{Arrival Time Data}
\label{sec: Arrival Time Data-2}

We need the linear operator \,$ \bfSigma $\, that is tangent to the 
forward operator \,$ \bfsigma $\, at some \,$ \bfg_0 $\,.
Formally,
\begin{equation}
\bfsigma(\bfg + \delta \bfg) \ = \ \bfsigma(\bfg) + \bfSigma \, \delta \bfg + 
O(\delta\bfg)^2 \quad . 
\end{equation}

It is easy to understand the meaning of \,$ \bfSigma $\,. While \,$ \bfsigma $\,
associates to any metric \,$ \bfg $\, some arrival times \,$ \sigma^i $\,,
the operator \,$ \bfSigma $\, associates to every metric perturbation
\,$ \delta \bfg $\, (around \,$ \bfg $\,) the perturbation 
\,$ \delta\sigma^i $\, of arrival times. Let us compute these perturbations.

\subsubsection{Hamiltonian Formulation of Finsler Geometry}

The Finsler space is a generalization of the Riemann space.
This generalization is appropriate for the description of
the propagation of light and many other waves.

Proper time $\tau$ in the Finsler space satisfies
the stationary Hamilton-Jacobi equation
\begin{equation}
H(x^\kappa,\tau_{,\mu}) \ = \ {\rm const.} \quad , 
\label{eq: klimes-201} 
\end{equation}
where $H(x^\kappa,p_\mu)$ is the Hamiltonian.
The geodesics can then be described by the Hamilton equations
\begin{equation} 
{{\rm d}x^\alpha\over{\rm d}\lambda}
 \ = \  {\partial H\over\partial p_\alpha}
\quad , 
\label{eq: klimes-202} 
\end{equation}
\begin{equation} {{\rm d}p_\alpha\over{\rm d}\lambda}
 \ = \ -{\partial H\over\partial x^\alpha}
\quad ,
\label{eq: klimes-203} \end{equation}
with initial conditions
\begin{equation} 
x^\alpha(\lambda_0) \ = \ x^\alpha_0
\qquad ; \qquad
 p_\alpha(\lambda_0) \ = \ \tau_{,\alpha}(x^\mu_0)
\quad . 
\label{eq: klimes-204} 
\end{equation}
Then
\begin{equation} \tau_{,\alpha}[x^\mu(\lambda)] \ = \ p_\alpha(\lambda)
\quad \label{eq: klimes-205} 
\end{equation}
along the geodesics.
Parameter $\lambda$ along a geodesic is determined by the form
of the Hamiltonian and by initial conditions (equation~\ref{eq: klimes-204}) 
for the geodesic.
Proper time $\tau$ along the geodesic is then given by
\begin{equation} \tau(\lambda) \ = \ \tau(\lambda_0)
 \ = \ \int_{\lambda_0}^{\lambda}{\rm d}\lambda\; p_\alpha
 {\partial H\over\partial p_\alpha}
\quad , \label{eq: klimes-206} \end{equation}
which follows from equations~\ref{eq: klimes-202} and~\ref{eq: klimes-205}.
Note that equal geodesics may be generated by various Hamiltonians.
For example, Hamiltonian
$\tilde{H}(x^\kappa,p_\mu)=F[H(x^\kappa,p_\mu)]$,
where $F(x)$ is an arbitrary function with a non--vanishing finite
derivative at $x$ equal to the right--hand side of equation~\ref{eq: klimes-201},
yields equal geodesics as Hamiltonian $H(x^\kappa,p_\mu)$.
The Hamiltonian is often chosen as a homogeneous function
of degree $N$ in $p_\alpha$.
Especially, homogeneous Hamiltonians of degrees $N=2$, $N=1$
or $N=-1$ are frequently used.

If the Hamiltonian is chosen as a homogeneous function of degree $N=2$
in $p_\alpha$, and is properly normalized, then
\begin{equation} g^{\alpha\beta}(x^\kappa,p_\mu)
 \ = \ {\partial^2 H\over\partial p_\alpha\partial p_\beta}
 (x^\kappa,p_\mu)
\quad 
\label{eq: klimes-207} \end{equation}
is the contravariant Finslerian metric tensor.
If metric tensor in equation~\ref{eq: klimes-207} is independent of $p_\mu$,
\begin{equation} 
g^{\alpha\beta}(x^\kappa,p_\mu) \ = \ g^{\alpha\beta}(x^\kappa)
\quad , 
\label{eq: klimes-208} 
\end{equation}
the Finsler space reduces to the {\it Riemann space}.

On the other hand, if we know the contravariant metric tensor,
we may construct a homogeneous Hamiltonian of degree $N$
in $p_\alpha$ as
\begin{equation} 
H(x^\kappa,p_\mu)
 \ = \ {1\over N}
 [p_\alpha g^{\alpha\beta}(x^\kappa,p_\mu)p_\beta]^{N\over2}
\quad . 
\label{eq: klimes-209} 
\end{equation}
Whereas degree $N$ may be arbitrary for spatial or time-like
geodesics, $N\ne2$ should be avoided for zero-length geodesics
in order to keep the right--hand sides of Hamilton 
equations~\ref{eq: klimes-202} and~\ref{eq: klimes-203} finite and 
non-vanishing identically.

For homogeneous Hamiltonians (equation~\ref{eq: klimes-209}), 
equation~\ref{eq: klimes-206} reads
\begin{equation} \tau(\lambda)
 \ = \ \tau(\lambda_0)
+\int_{\lambda_0}^{\lambda}{\rm d}\lambda\;
 [p_\alpha g^{\alpha\beta}(x^\kappa,p_\mu)p_\beta]^{N\over2}
\quad , \label{eq: klimes-210} \end{equation}
and equation~\ref{eq: klimes-202} yields
\begin{equation} 
{{\rm d}x^\alpha\over{\rm d}\lambda} g_{\alpha\beta}
 {{\rm d}x^\beta\over{\rm d}\lambda}
 \ = \ [p_\alpha g^{\alpha\beta}p_\beta]^{N-1}
\quad . 
\label{eq: klimes-211} 
\end{equation}
Considering equation~\ref{eq: klimes-211}, equation~\ref{eq: klimes-210}, 
can be expressed in the form
\begin{equation} 
\tau(\lambda)
 \ = \ \tau(\lambda_0)
+\int_{\lambda_0}^{\lambda}{\rm d}\lambda\;
 \left[{{\rm d}x^\alpha\over{\rm d}\lambda}
 g_{\alpha\beta}(x^\kappa,p_\mu)
 {{\rm d}x^\beta\over{\rm d}\lambda}\right]^{N\over2(N-1)}
\quad . 
\label{eq: klimes-212} 
\end{equation}
In the Hamiltonian formulation, the Finsler geometry is no more
complex than the Riemann geometry.

\subsubsection{Perturbation of Proper Time}

The first-order perturbation of proper time (equation~\ref{eq: klimes-206}) is
(Klime\v{s}, 2002, eq.~25)
\begin{equation} 
\delta\tau(\lambda) \ = \ \delta\tau(\lambda_0)
-\int_{\lambda_0}^{\lambda}{\rm d}\lambda\;\delta H
\quad . 
\label{eq: klimes-213}
 \end{equation}
If we wish to perform perturbations with respet to
the components of the metric tensor along zero--length
space--time geodesics, homogeneous Hamiltonians
(equation~\ref{eq: klimes-209}) should be of degree $N=2$ to avoid zero
or infinite perturbations $\delta H$ of the Hamiltonian.

One alternative to the present Hamiltomian formulation, 
would be to use a Lagrangian formulation of the first degree,
this leading to the usual Fermat's integral.
There are four reasons why the formulation here presented is better:
\begin{itemize}
\item
Perturbations of a homogeneous Lagrangian of degree $N$
with respect to the components of the metric tensor
are zero for $N>2$ and infinite for $N<2$,
which results in singularities in the computation;

\item
Hamilton equations beak down for $N\ne2$,
which would prevent us from using efficient tools
of Hamiltonian formulation;

\item
Dual Legendre transform between homogeneous Hamiltonian
and Lagrangian of the first degree is not possible
(\v{C}erven\'{y}, 2002), which also holds for
spatial and time--like geodesics;

\item
The integral is generally complex-valued for indefinite
metric tensors.
\end{itemize}

In the following, we shall thus consider an homogeneous Hamiltonian
(equation~\ref{eq: klimes-209}) of degree $N=2$,
\begin{equation} 
H(x^\kappa,p_\mu)
 \ = \ {1\over 2}
 p_\alpha\,g^{\alpha\beta}(x^\kappa,p_\mu)\,p_\beta
\quad . 
\label{eq: klimes-214} 
\end{equation}
Equation~\ref{eq: klimes-202} then reads
\begin{equation} 
{{\rm d}x^\alpha\over{\rm d}\lambda}
 \ = \  g^{\alpha\beta}(x^\kappa,p_\mu)p_\beta
\quad ,
 \label{eq: klimes-215} 
 \end{equation}
and equation~\ref{eq: klimes-213}, with
\,$ \delta\tau(\lambda_0) = 0 $\,,
reads
\begin{equation}
\delta\tau(\lambda)
 \ = \ -{1\over 2}
 \int_{\lambda_0}^{\lambda}{\rm d}\lambda\;
 p_\alpha\delta g^{\alpha\beta}p_\beta
\quad . \label{eq: klimes-217}
\end{equation}
Inserting \,$ \delta g^{\alpha\beta}
 = -g^{\alpha\kappa} \delta g_{\kappa\mu} g^{\mu\beta} $\,, 
we obtain
\begin{equation} 
\delta\tau(\lambda)
 \ = \ {1\over 2}
 \int_{\lambda_0}^{\lambda}{\rm d}\lambda\;
 p_\alpha
 g^{\alpha\kappa} \delta g_{\kappa\mu} g^{\mu\beta}
 p_\beta \quad . 
\label{eq: klimes-219}
 \end{equation}
Inserting equation~\ref{eq: klimes-215} into equation~\ref{eq: klimes-219}, 
we arrive at
\begin{equation} 
\delta\tau(\lambda)
 \ = \ {1\over 2}
 \int_{\lambda_0}^{\lambda}{\rm d}\lambda\;
 {{\rm d}x^\alpha\over{\rm d}\lambda}
 \delta g_{\alpha\beta}
 {{\rm d}x^\beta\over{\rm d}\lambda}
\quad . \label{eq: klimes-220} 
\end{equation}

\subsubsection{Perturbation of arrival time}

Assume the trajectory
\begin{equation} 
x^i  \ = \  y^i(\sigma)
 \quad \label{eq: klimes-221} 
 \end{equation}
parametrized by proper time $\sigma$ along it
(in general, $\sigma$ may represent an arbitrary parameter
along the trajectory).
A light signal emitted at the given point will hit
the given trajectory at proper time $\sigma=\sigma^0$.
Assume now that the space--time metric is perturbed from
$g_{ij}$ to $g_{ij}+\delta g_{ij}$.The light signal will
now hit the trajectory at proper time $\sigma^0+\delta\sigma$.
We shall now derive the first order relation between
$\delta\sigma$ and delta $g_{ij}$.

The space-time wavefront may be expressed in the form
\begin{equation} 
\tau(x^\alpha)  \ = \  0
 \quad , 
 \label{eq: klimes-222} 
 \end{equation}
where $\tau(x^\alpha)$ is measured along the geodesic from the given point
to point $x^\alpha$. The geodesics can be calculated by Hamiltonian
ray tracing from the given point.

Proper time $\sigma$ at the point of intersection of the trajectory
with the space-time wavefront then satisfies equation
\begin{equation} 
\tau(y^\alpha(\sigma))  \ = \  0
 \quad . 
 \label{eq: klimes-223} 
 \end{equation}
Perturbation of this equation yields
\begin{equation} 
\delta \tau(y^\beta(\sigma))
+\tau_{,\alpha}(y^\beta(\sigma))\;
 {{\rm d}y^\alpha\over{\rm d}\sigma}(\sigma)\;\delta\sigma  \ = \  0
 \quad . 
 \label{eq: klimes-224} 
 \end{equation}
Then
\begin{equation} 
\delta\sigma  \ = \  - { \delta \tau(y^\beta(\sigma)) \over
 \tau_{,\alpha}(y^\beta(\sigma))\;
 {{\rm d}y^\alpha\over{\rm d}\sigma}(\sigma) }
 \quad . 
 \label{eq: klimes-225}
  \end{equation}
Inserting $p_\alpha$ from equation~\ref{eq: klimes-215} for $\tau_{,\alpha}$ and
equation~\ref{eq: klimes-220} for $\delta \tau(y^\beta(\sigma))$,
equation~\ref{eq: klimes-225} can be expressed in the form
\begin{equation} 
\delta\sigma
 \ = \ -{1\over 2}
\left[
 {{\rm d}y^\beta\over{\rm d}\sigma}\,g_{\alpha\beta}\,
 {{\rm d}x^\alpha\over{\rm d}\lambda}
\right]^{-1}
\int_{\lambda_0}^{\lambda}{\rm d}\lambda\;
{{\rm d}x^\alpha\over{\rm d}\lambda}
\delta g_{\alpha\beta}
{{\rm d}x^\beta\over{\rm d}\lambda}
 \quad . 
 \label{eq: klimes-226}
 \end{equation}


\subsection{A Priori Information on the Metric}
\label{sec: Distance Between two Metrics}

Let \,$ \bfg_\prior $\, some reference space-time metric (for instance the
Minkowski or the Schwarzschild metric), and let \,$ \bfg $\, be the
actual metric. 
In the simple (and a little bit simplistic) approach proposed here,
it is assumed that the difference
\begin{equation}
\bfg - \bfg_\prior 
\label{eq: dii-mert}
\end{equation}
is small, and is assumed to be a random realization of a Gaussian random 
field with zero mean and prescribed covariance. Because in the light
coordinates used here it is the contravariant metric that has some simple
properties, the difference in equation~\ref{eq: dii-mert} is taken using
the contravariant components.

To obtain a reasonable model of covariance operator for the metric, 
we could perform a thought experiment. We imagine a large number of metric 
fields, all of the form
\begin{equation}
\{ g^{\alpha\beta} \} \ = \ 
\begin{pmatrix}
0 & g^{12} & g^{13} & g^{14} \\
g^{12} & 0 & g^{23} & g^{24} \\
g^{13} & g^{23} & 0 & g^{34} \\
g^{14} & g^{24} & g^{34} & 0 \\
\end{pmatrix} 
\end{equation}
at every point, all smoothly varying over space-time, and with the quantities
\,$ g^{\alpha\beta} $\, randomly varying around the values corresponding
to the reference metric, with prescribed, simple probability distributions 
(independent, to start with). The we could evaluate the covariance of such a 
`random field' using the direct definition of covariance:
\begin{equation}
\begin{split}
& C^{\alpha\beta\mu\nu}
(\tau^1,\tau^2,\tau^3,\tau^4,\sigma^1,\sigma^2,\sigma^3,\sigma^4) \ = \ \\[5pt]
& \overline{ \left(g^{\alpha\beta}(\tau^1,\tau^2,\tau^3,\tau^4) - 
\overline{g^{\alpha\beta}(\tau^1,\tau^2,\tau^3,\tau^4)}\right) \ 
\left(g^{\mu\nu}(\sigma^1,\sigma^2,\sigma^3,\sigma^4) - 
\overline{g^{\mu\nu}(\sigma^1,\sigma^2,\sigma^3,\sigma^4)}\right) } \ ,
\end{split}
\end{equation}
where \,$ \overline{x} $\, means the mean value of \,$ x $\,. 
The mean metric \,$ \overline{g^{\alpha\beta}} $\, would be the reference
metric.

Another option is to try to insert more constraints that we know are satisfied
by the metric. For instance, Pozo (2005) shows that the metric has necessarily
the form
\begin{equation}
\begin{pmatrix}
0 & g^{12} & g^{13} & g^{14} \\
g^{12} & 0 & g^{23} & g^{24} \\
g^{13} & g^{23} & 0 & g^{34} \\
g^{14} & g^{24} & g^{34} & 0 \\
\end{pmatrix} \ = \ 
\begin{pmatrix}
a & 0 & 0 & 0 \\ 0 & b & 0 & 0 \\ 0 & 0 & c & 0 \\ 0 & 0 & 0 & d \\ 
\end{pmatrix} \
\begin{pmatrix}
0 & A & B & 1 \\
A & 0 & 1 & B \\
B & 1 & 0 & A \\
1 & B & A & 0 \\
\end{pmatrix} \
\begin{pmatrix}
a & 0 & 0 & 0 \\ 0 & b & 0 & 0 \\ 0 & 0 & c & 0 \\ 0 & 0 & 0 & d \\ 
\end{pmatrix} \quad ,
\end{equation}
where the constants \,$ \{a,b,c,d\} $\, are positive, and the constants
\,$ \{A,B $\, should satisfy the constraint that a triangle exists
in the Euclidean plane whose sides have the lengths \,$ \{A,B,1\} $\,.
One could perhaps use the six quantities \,$ \{a,b,c,d,A,B\} $\, as
basic quantities, and assume a Gaussian distribution for some simple
functions of them.

We do not explore yet this possibility.
Also, it is very likely that the basic variable to be used in the optimization
problem is not the metric \,$ g^{\alpha\beta} $\,,
but the logarithmic metric. This point is, for the time being, 
not examined. 

We don't try to be more specific at this point, we simply assume that some
covariance function
\begin{equation}
C^{\alpha\beta\mu\nu}
(\tau^1,\tau^2,\tau^3,\tau^4,\sigma^1,\sigma^2,\sigma^3,\sigma^4)
\end{equation}
is chosen.
The inverse \,$ \bfW = \bfC^{-1} $\, of the covariance operator 
(a distribution) has the kernel
\begin{equation}
W_{\alpha\beta\mu\nu}(\tau^1,\tau^2,\tau^3,\tau^4;
\sigma^1,\sigma^2,\sigma^3,\sigma^4)
\quad .
\end{equation}
By definition (formally)
\begin{equation}
\begin{split}
\int dv(\rho^1,\rho^2,\rho^3,\rho^4) \ 
& W_{\alpha\beta\rho\sigma}
(\tau^1,\tau^2,\tau^3,\tau^4;\rho^1,\rho^2,\rho^3,\rho^4) \times \\
& \times C^{\rho\sigma\mu\nu}(\rho^1,\rho^2,\rho^3,\rho^4;\sigma^1,\sigma^2,\sigma^3,\sigma^4) \ = \\[6 pt]
& = \ \delta_\alpha^\mu \, \delta_\beta^\nu \,
\delta(\tau^1-\sigma^1) \, \delta(\tau^2-\sigma^2) \, \delta(\tau^3-\sigma^3) \, \delta(\tau^4-\sigma^4) \quad ,
\end{split}
\end{equation}
where
\begin{equation}
dv(\rho^1,\rho^2,\rho^3,\rho^4) \ = \ 
\sqrt{ - \det \bfg_\prior(\rho^1,\rho^2,\rho^3,\rho^4) } \ 
d\rho^1 \, d\rho^2 \, d\rho^3 \, d\rho^4 \quad .
\end{equation}

The operators \,$ \bfC(\bfg) $\, and \,$ \bfW(\bfg) $\, being symmetric 
and positive definite, define a bijection between \,$ \calG $\,,
the space of metric field perturbations and its dual, 
\,$ \calG^\ast $\,.
We shall write
\begin{equation}
\delta\hat{\bfg} \ = \ \bfW \, \delta \bfg 
\qquad ; \qquad 
\delta \bfg \ = \ \bfC \, \delta \hat{\bfg} \quad .
\end{equation}
Explicitly,
\begin{equation}
\begin{split}
\delta \hat{g}_{\alpha\beta}(\tau^1,\tau^2,\tau^3,\tau^4) \ & = \ 
\int dv(\rho^1,\rho^2,\rho^3,\rho^4) \\
& W_{\alpha\beta\mu\nu}
(\tau^1,\tau^2,\tau^3,\tau^4;\sigma^1,\sigma^2,\sigma^3,\sigma^4) \ 
\delta g^{\mu\nu}(\sigma^1,\sigma^2,\sigma^3,\sigma^4) 
\end{split}
\end{equation}
and
\begin{equation}
\begin{split}
\delta g^{\alpha\beta}(\tau^1,\tau^2,\tau^3,\tau^4) \ & = \ 
\int dv(\rho^1,\rho^2,\rho^3,\rho^4) \\
& C^{\alpha\beta\mu\nu}
(\tau^1,\tau^2,\tau^3,\tau^4;\sigma^1,\sigma^2,\sigma^3,\sigma^4) \ 
\delta\hat{g}_{\mu\nu}(\sigma^1,\sigma^2,\sigma^3,\sigma^4) \quad .
\end{split}
\label{eq: notexc}
\end{equation}

The duality product of a dual metric field perturbation \,$ \delta\hat{\bfg} $\, 
by a metric field perturbation \,$ \delta\bfgamma $\, is defined as
\begin{equation}
\langle \, \delta\hat{\bfg} \, , \, \delta\bfgamma \, \rangle \ = \ 
\int dv(\tau^1,\tau^2,\tau^3,\tau^4) \ 
\delta\hat{g}_{\alpha\beta}(\tau^1,\tau^2,\tau^3,\tau^4) \
\delta\gamma^{\alpha\beta}(\tau^1,\tau^2,\tau^3,\tau^4) \quad , 
\end{equation}
the scalar product of two metric field perturbations is
\begin{equation}
\delta \bfg_1 \cdot \delta \bfg_2 \ = \ 
\langle \, \bfW \, \delta\bfg_1 \, , \, \delta\bfg_2 \, \rangle
\quad ,
\end{equation}
and the norm of a metric field perturbation is
\begin{equation}
\Vert \, \delta\bfg \, \Vert_{\bfC_\bfg} \ = \ \sqrt{ \, \delta\bfg \cdot \delta\bfg \, } \quad .
\end{equation}

Denoting by \,$ \bfg_\prior $\, the a priori metric and by \,$ \bfg $\, our estimation 
of the actual metric field, we are later going to impose that the squared norm
\begin{equation}
2 \, S_{\bfg}(\bfg) \ = \ \Vert \, \bfg - \bfg_\prior \, \Vert^2_{\bfC_\bfg}
\label{eq: el-ese-del-C}
\end{equation}
is small.


\subsection{Newton Algorithm}
\label{sec: Newton Algorithm}

While in section~\ref{sec: Iterative Algorithm} we have examined the simple
steepest descent algorithm, let us now develop the quasi-Newton method. 
To obtain the actual algorithm, one may use the formulas developed in 
Tarantola (2005). The resulting iterative algorithm can be written
\begin{equation}
\bfg_{k+1} \ = \ \bfg_k - \bfH_k^{-1} \, \bfgamma_k \quad ,
\label{eq: Newton-algorithm}
\end{equation}
where the `Hessian operator' \,$ \bfH_k $\, is
\begin{equation}
\begin{split}
\bfH_k \ = \ \bfI 
& + (\bfZ_k \, \bfC_\bfg)^t \, \bfC_{\bfz}^{-1} \, \bfZ_k 
+ (\bfT_k \, \bfC_\bfg)^t \, \bfC_{\bft}^{-1} \, \bfT_k 
+ (\bfSigma_k \, \bfC_\bfg)^t \, \bfC_{\bfsigma}^{-1} \, \bfSigma_k \\
& + (\bfA_k \, \bfC_\bfg)^t \, \bfC_{\bfa}^{-1} \, \bfA_k 
+ (\bfPi_k \, \bfC_\bfg)^t \, \bfC_{\bfpi}^{-1} \, \bfPi_k 
+ (\bfOmega_k \, \bfC_\bfg)^t \, \bfC_{\bfomega}^{-1} \, \bfOmega_k \quad ,
\end{split}
\label{eq: Hessian operator}
\end{equation}
the `gradient vector' is
\begin{equation}
\begin{split}
\bfgamma_k \ = \ (\bfg_k - \bfg_\prior) 
& + (\bfZ_k \, \bfC_\bfg)^t \, \bfC_\bfz^{-1} \, (\bfz(\bfg_k) - \bfone) \\[3 pt]
& + (\bfT_k \, \bfC_\bfg)^t \, \bfC_\bft^{-1} \, (\bft(\bfg_k) - \bft_\obs) \\[3 pt]
& + (\bfSigma_k \, \bfC_\bfg)^t \, \bfC_\bfsigma^{-1} \, (\bfsigma(\bfg_k) - \bfsigma_\obs) \\[3 pt]
& + (\bfA_k \, \bfC_\bfg)^t \, \bfC_\bfa^{-1} \, (\bfa(\bfg_k) - \bfa_\obs) \\[3 pt]
& + (\bfPi_k \, \bfC_\bfg)^t \, \bfC_\bfpi^{-1} \, (\bfpi(\bfg_k) - \bfpi_\obs) \\[3 pt]
& + (\bfOmega_k \, \bfC_\bfg)^t \, \bfC_\bfomega^{-1} \, (\bfomega(\bfg_k) - \bfomega_\obs) \quad , \\
\end{split}
\label{eq: gradient vector}
\end{equation}
where the linear operators \,$ \bfZ_k $\,, \,$ \bfT_k $\,, \,$ \bfSigma_k $\,, 
\,$ \bfA_k $\,, \,$ \bfPi_k $\,, and \,$ \bfOmega_k $\,, are the Fr\'echet
derivatives (tangent linear applications) of the operators 
\,$ \bfz(\bfg) $\,, \,$ \bft(\bfg) $\,, \,$ \bfsigma(\bfg) $\,, 
\,$ \bfa(\bfg) $\,, \,$ \bfpi(\bfg) $\,, and \,$ \bfomega(\bfg) $\,, 
introduced in equations~\ref{eq: defino-t}, \ref{eq: defino-z},
\ref{eq: defino-sigma}, \ref{eq: defino-a}, \ref{eq: defino-s}, 
and~\ref{eq: defino-omega}, all the operators evaluated for \,$ \bfg = \bfg_k $\,, 
and where \,$ \bfL^t $\, denotes the transpose of a linear operator \,$ \bfL $\,.
We say \emph{transpose} operators, better than \emph{dual}
operators, because the difference between the two notions matters inside the
theory of least-squares.

All the linear operators just introduced are evaluated in 
section~\ref{sec: Transpose of a Linear Application}.
But before going into these details, some comments on the
iterative algorithm are needed. 

The quasi-Newton algorithm~\ref{eq: Newton-algorithm} can be initialized at
an arbitrary point (i.e., at any metric field) \,$ \bfg_0 $\,. If working 
in the vicinity of an ordinary planet, the present problem will only be
mildly nonlinear, and the convergence point will be independent of the
initial point. The simplest choice, of course, is
\begin{equation}
\bfg_0 \ = \ \bfg_\prior \quad .
\end{equation}

Before entering on the problem of how many iterations must
the done in practice, let us take the strict mathematical point of view that,
in principle, an infinite number of iterations should be performed. 
The optimal estimate of the space-time metric would then be
\begin{equation}
\widetilde{\bfg} \ = \ \bfg_\infty \quad .
\label{eq: infinity-1}
\end{equation}
The least-squared method not only provides an optimal solution, it also 
provides a mean of estimating the uncertainties on this solution. 
It can be shown (Tarantola, 2005) that these uncertainties are those
represented by the covariance operator
\begin{equation}
\boxed{ \qquad 
\widetilde{\bfC}_\bfg \ = \ \bfH_\infty^{-1} \, \bfC_\bfg 
\quad . \quad } 
\label{eq: infinity-2}
\end{equation}
Crudely speaking, we started with the a priori metric \,$ \bfg_\prior $\,,
with uncertainties represented by the covariance operator \,$ \bfC_\bfg $\,, 
and we end up with the a posteriori metric \,$ \widetilde{\bfg} $\,,
with uncertainties represented by the covariance operator 
\,$ \widetilde{\bfC}_\bfg $\,.

The practical experience we have with the quasi-Newton algorithm for travel-time
fitting problems suggests that the algorithm should converge to the proper
solution (with sufficient accuracy) in a few iterations (less than 10).
Then, for all practical purposes, we can replace \,$ \infty $\, by \,$ 10 $\,  
in the two equations~\ref{eq: infinity-1}--\ref{eq: infinity-2}.

An important practical consideration is the following. The Hessian operator
(equation~\ref{eq: Hessian operator}) shall be completely characterized below, 
and the different covariance operators shall be directly given. But the algorithm
in equations~\ref{eq: Newton-algorithm}--\ref{eq: gradient vector} contains the
inverse of these linear operators. 
It is a very basic result of numerical analysis (Ciarlet, 1982) that the numerical resolution of a linear system may be dramatically more economical than the numerical
evaluation of the inverse of a linear operator.
Therefore, we need to rewrite the quasi-Newton algorithm replacing every
occurrence of the inverse of an operator by the associated resolution of a linear
system.

Let us start by the evaluation of the gradient vector \,$ \bfgamma_k $\,.
Expression~\ref{eq: gradient vector} can be rewritten
\begin{equation}
\boxed{ \qquad
\begin{split}
\bfgamma_k \ = \ \delta\bfg_k
& + (\bfZ_k \, \bfC_\bfg)^t \, \delta\bfz_k^\ast  
+ (\bfT_k \, \bfC_\bfg)^t \, \delta\bft_k^\ast  
+ (\bfSigma_k \, \bfC_\bfg)^t \, \delta\bfsigma_k^\ast \\
& + (\bfA_k \, \bfC_\bfg)^t \, \delta\bfa_k^\ast 
+ (\bfPi_k \, \bfC_\bfg)^t \, \delta\bfpi_k^\ast 
+ (\bfOmega_k \, \bfC_\bfg)^t \, \delta\bfomega_k^\ast \quad , \\ 
\end{split}
\quad } 
\label{eq: gradient vector 2}
\end{equation}
where
\begin{equation}
\delta\bfg_k \ = \ \bfg_k - \bfg_\prior \quad ,
\end{equation}
and where the vectors
\,$ \delta\bfz_k^\ast $\,, \,$ \delta\bft_k^\ast $\,, 
\,$ \delta\bfsigma_k^\ast $\, \,$ \delta\bfa_k^\ast $\,, \,$ \delta\bfpi_k^\ast $\,, 
and \,$ \delta\bfomega_k^\ast $\,, are the respective solutions of the linear systems
\begin{equation}
\boxed{ \qquad
\begin{split}
\bfC_\bfz \, \delta\bfz_k^\ast \ & = \ \bfz(\bfg_k) - \bfone \\[3 pt]
\bfC_\bft \, \delta\bft_k^\ast \ & = \ \bft(\bfg_k) - \bft_\obs \\[3 pt]
\bfC_\bfsigma \, \delta\bfsigma_k^\ast \ & = \ \bfsigma(\bfg_k) - \bfsigma_\obs \\[3 pt]
\bfC_\bfa \, \delta\bfa_k^\ast \ & = \ \bfa(\bfg_k) - \bfa_\obs \\[3 pt]
\bfC_\bfpi \, \delta\bfpi_k^\ast \ & = \ \bfpi(\bfg_k) - \bfpi_\obs \\[3 pt]
\bfC_\bfomega \, \delta\bfomega_k^\ast \ & = \ \bfomega(\bfg_k) - \bfomega_\obs \quad . \\
\end{split}
\quad } 
\label{eq: gradient vector 3}
\end{equation}

Once the gradient vector \,$ \bfgamma_k $\, is evaluated, one can turn to
the iterative step (equation~\ref{eq: Newton-algorithm}). It can be written
\begin{equation}
\boxed{ \qquad 
\bfg_{k+1} \ = \ \bfg_k - \Delta\bfg_k \quad , \quad } 
\end{equation}
where \,$ \Delta\bfg_k $\, is the solution of the linear system
\begin{equation}
\bfH_k \, \Delta\bfg_k \ = \ \bfgamma_k \quad .
\end{equation}
Using the expression~\ref{eq: Hessian operator} for the operator \,$ \bfH_k $\, 
we can equivalently say that \,$ \Delta\bfg_k $\, is the solution of the linear system
\begin{equation}
\boxed{ \qquad
\begin{split}
\Delta\bfg_k
& + (\bfZ_k \, \bfC_\bfg)^t \, \Delta\bfz^\ast_k
+ (\bfT_k \, \bfC_\bfg)^t \, \Delta\bft^\ast_k
+ (\bfSigma_k \, \bfC_\bfg)^t \, \Delta\bfsigma^\ast_k \\
& + (\bfA_k \, \bfC_\bfg)^t \, \Delta\bfa^\ast_k 
+ (\bfPi_k \, \bfC_\bfg)^t \, \Delta\bfpi^\ast_k 
+ (\bfOmega_k \, \bfC_\bfg)^t \, \Delta\bfomega^\ast_k \ = \ \bfgamma_k 
\quad , 
\end{split}
\quad } 
\label{eq: final linear system}
\end{equation}
where, when introducing the vectors
\begin{equation}
\boxed{ \qquad
\begin{split}
\Delta\bfz_k \ & = \ \bfZ_k \, \Delta\bfg_k \\
\Delta\bft_k \ & = \ \bfT_k \, \Delta\bfg_k \\
\Delta\bfsigma_k \ & = \ \bfSigma_k \, \Delta\bfg_k \\
\Delta\bfa_k \ & = \ \bfA_k \, \Delta\bfg_k \\
\Delta\bfpi_k \ & = \ \bfPi_k \, \Delta\bfg_k \\
\Delta\bfomega_k \ & = \ \bfOmega_k  \, \Delta\bfg_k 
\quad , \\ 
\end{split}
\quad } 
\label{eq: intermediary vectors}
\end{equation}
the vectors \,$ \Delta\bfz^\ast_k $\,, 
\,$ \Delta\bft^\ast_k $\,, \,$ \Delta\bfsigma^\ast_k $\,, \,$ \Delta\bfa^\ast_k $\,, 
\,$ \Delta\bfpi^\ast_k $\,, and \,$ \Delta\bfomega^\ast_k $\,, 
are the respective solutions of the linear systems
\begin{equation}
\boxed{ \qquad
\begin{split}
\bfC_{\bfz} \, \Delta\bfz^\ast_k \ & = \ \Delta\bfz_k \\
\bfC_{\bft} \, \Delta\bft^\ast_k \ & = \ \Delta\bft_k \\
\bfC_{\bfsigma} \, \Delta\bfsigma^\ast_k \ & = \ \Delta\bfsigma_k \\
\bfC_{\bfa} \, \Delta\bfa^\ast_k \ & = \ \Delta\bfa_k \\
\bfC_{\bfpi} \, \Delta\bfpi^\ast_k \ & = \ \Delta\bfpi_k \\
\bfC_{\bfomega} \, \Delta\bfomega^\ast_k \ & = \ \Delta\bfomega_k \quad . \\
\end{split}
\quad } 
\label{eq: final weighting}
\end{equation}

In equations~\ref{eq: gradient vector 2} and~\ref{eq: final linear system}
one needs to evaluate vectors whose generic form is
\begin{equation}
\bfb \ = \ (\bfL \, \bfC)^t \, \bfa \quad .
\end{equation}
the vector \,$ \bfa $\, being known.
They involve the transpose of an operator. To evaluate these vectors one must
resort to the very definition of transpose operator.
By definition, the operator \,$ (\bfL \, \bfC)^t $\, is the transpose of
the linear operator \,$ (\bfL \, \bfC) $\, if, and only if, for any 
\,$ \bfa^\ast $\, and any
\,$ \bfa $\,,
\begin{equation}
\langle \, \bfa^\ast \, , \, (\bfL \, \bfC)^t \, \bfa \, \rangle \ = \ 
\langle \, (\bfL \, \bfC) \, \bfa^\ast \, , \, \bfa \, \rangle \quad .
\end{equation}
As the linear tangent operators are characterized (for all the nonlinear
applications appearing above), we know how to write the right-hand 
side of this equation explicitly. As the vector \,$ \bfa $\, is known,
the condition that the obtained expression must hold for any vector
\,$ \bfa^\ast $\, gives an explicit expression for 
\,$ \bfb \ = \ (\bfL \, \bfC)^t \, \bfa $\,.
Appendixes~\ref{sec: Appendix Einstein Equation}
and~\ref{sec: Arrival Time Data} provide two examples of this kind of evaluations.


\subsection{Kalman Filter}
\label{sec: Kalman Filter}

Assume that some \emph{linear model} allows to make a \emph{preliminary} 
prediction of the state of the system at time \,$ k $\, in terms of the state 
of the system at time \,$ k-1 $\,
(we retain here the notations in Grewal et al.~(2001)):
\begin{equation}
x_k^- \ = \ \Phi_k \ x_{k-1}^+ \quad .
\label{eq: Kalman-1}
\end{equation}
If the uncertainties we had on \,$ x_{k-1}^+ $\, are represented by the
covariance matrix \,$ P_{k-1}^+ $\, and if the prediction by the linear model 
\,$ \Phi_k $\, has uncertainties described by the covariance matrix \,$ Q_{k-1}$\,,
the uncertainty we have on \,$ x_k^- $\, is represented by the covariance matrix
\begin{equation}
P_k^- \ = \ \Phi_k \ P_{k-1}^+ \ \Phi_k^t + Q_{k-1} \quad .
\label{eq: Kalman-2}
\end{equation}

So we have the prior value \,$ x_k^- $ \, with uncertainties described by the
prior covariance matrix \,$ P_k^- $\,. 
To pass from the preliminary estimate \,$ x_k^- $\, to the actual estimate
\,$ x_k^+ $\, we now use some observed data \,$ z_k $\, that is assumed to be
related to \,$ x_k^+ $\, via a linear relation \,$ z_k \approx H_k \, x_k^+ $\,,
with uncertainties described by the covariance matrix \,$ R_k $\,. The standard
theory of linear least-squares then provides the posterior estimate as
\begin{equation}
x_k^+ \ = \ x_k^- + P_k^- \ H_k^t \ ( \, H_k \ P_k^- \ H_k^t + R_k \, )^{-1} \ 
( \, z_k - H_k \ x_k^- \, ) \quad , 
\label{eq: Kalman-3}
\end{equation}
that has uncertainties represented by the posterior covariance matrix
\begin{equation}
P_k^+ \ = \ P_k^- - P_k^- \ H_k^t \ 
( \, H_k \ P_k^- \ H_k^t + R_k \, )^{-1} \ H_k \ P_k^- \quad .
\label{eq: Kalman-4}
\end{equation}

Then, if at each time step we input \,$ \Phi_k $\,, \,$ Q_{k-1} $\,, 
\,$ z_k $\,, \,$ H_k $\,, and \,$ R_k $\,, 
equations~\ref{eq: Kalman-1}--\ref{eq: Kalman-4} allow to have a continuous
estimation of the state of the system, \,$ x_k^+ $\,, together with an estimation
of the uncertainties, \,$ P_k^+ $\,.

The reader may recognize that 
equations~\ref{eq: Kalman-3}--\ref{eq: Kalman-4} are identical to the 
standard equations of linear least-squares theory
(see equations~3.37 and 3.38 in Tarantola (2005)). The matrix 
\begin{equation}
K_k \ = \ P_k^- \ H_k^t \ ( \, H_k \ P_k^- \ H_k^t + R_k \, )^{-1} \quad ,
\end{equation}
that appears in the two equations~\ref{eq: Kalman-3}--\ref{eq: Kalman-4},
is called the `Kalman gain matrix'.

\emph{Example.} As a simple example, consider, in non-relativistic physics, 
the trajectory of a mass that has been equipped with some sensors. We can choose
to represent the state of the system at any time by a 9-dimensional vector \,$ x $\,, 
that contains the three values of the position, the three values of the velocity
and the three values of the acceleration. Assume that, as a result of the previous 
iteration, at some moment we have the estimation \,$ x_{k-1}^+ $\, with 
uncertainties \,$ P_{k-1}^+ $\,. 
Equation~\ref{eq: Kalman-1} may simply correspond to the use of the velocity to
extrapolate the position one step in time, to use the acceleration to extrapolate
the velocity, and to keep the acceleration unchanged. This perfectly characterizes
the matrix \,$ \Phi_k $\,. Equation~\ref{eq: Kalman-2} then is used to update
the estimation of uncertainty, where we can take for \,$ Q_{k-1} $\, something
as simple as a zero matrix excepted for the three diagonal elements associated
to the acceleration, where a small variance will take into account that our
extrapolation of acceleration is uncertain. The data \,$ z_k $\, may consist in the 
output of some sensors, like accelerometers or data from a satellite positioning
system. The relation
\,$ z_k = H_k \, x_k^- $\, would correspond to the theoretical calculation of
the data \,$ z_k $\, given the state \,$ x_k^- $\,. This is not a linear relation, 
and the theory should be developed to directly account for this, but if the time steps
are small enough, we can always linearize the theory, this then defining the
matrix \,$ H_k $\,. Denoting now by \,$ z_k $\, the actual output of the sensors,
and by \,$ R_k $\, the experimental uncertainties, equation~\ref{eq: Kalman-3}
is used to obtain our second estimation of the state of the system, 
equation~\ref{eq: Kalman-4} providing the associated uncertainties.

\end{document}